\documentclass[floatfix,superscriptaddress,longbibliography,pre]{revtex4-1}
\usepackage{bbm,amsmath,amsfonts,graphicx, mathrsfs, float, mathtools, cases}
\usepackage{hyperref}
\newcommand{\1}{\mathbbm 1}

\newcommand{\Bm}{\mathbf m}
\newcommand{\Bs}{\mathbf s}
\newcommand{\dd}{\partial}
\newcommand{\sm}{\setminus}

\begin{document}
\title{Containing epidemic outbreaks by message-passing techniques}
\author{F.~Altarelli}
\affiliation{Department of Applied Science and Technology, Politecnico di Torino, Corso Duca degli Abruzzi 24, 10129 Torino, Italy}
\affiliation{Collegio Carlo Alberto, Via Real Collegio 30, 10024 Moncalieri, Italy}
\author{A.~Braunstein}
\affiliation{Department of Applied Science and Technology, Politecnico di Torino, Corso Duca degli Abruzzi 24, 10129 Torino, Italy}
\affiliation{Human Genetics Foundation, Via Nizza 52, 10126 Torino, Italy}
\affiliation{Collegio Carlo Alberto, Via Real Collegio 30, 10024 Moncalieri, Italy}
\author{L.~Dall'Asta}
\affiliation{Department of Applied Science and Technology, Politecnico di Torino, Corso Duca degli Abruzzi 24, 10129 Torino, Italy}
\affiliation{Collegio Carlo Alberto, Via Real Collegio 30, 10024 Moncalieri, Italy}
\author{J.R.~Wakeling}
\affiliation{Department of Applied Science and Technology, Politecnico di Torino, Corso Duca degli Abruzzi 24, 10129 Torino, Italy}
\author{R.~Zecchina}
\affiliation{Department of Applied Science and Technology, Politecnico di Torino, Corso Duca degli Abruzzi 24, 10129 Torino, Italy}
\affiliation{Human Genetics Foundation, Via Nizza 52, 10126 Torino, Italy}
\affiliation{Collegio Carlo Alberto, Via Real Collegio 30, 10024 Moncalieri, Italy}

\begin{abstract}
	The problem of targeted network immunization can be defined as the one of finding a subset of nodes in a network to immunize or vaccinate in order to minimize a tradeoff between the cost of vaccination and the final (stationary) expected infection under a given epidemic model. Although computing the expected infection is a hard computational problem, simple and efficient mean-field approximations have been put forward in the literature in recent years. The optimization problem can be recast into a constrained one in which the constraints enforce local mean-field equations describing the average stationary state of the epidemic process. For a wide class of epidemic models, including the susceptible-infected-removed and the susceptible-infected-susceptible models, we define a message-passing approach to network immunization that allows us to study the statistical properties of epidemic outbreaks in the presence of immunized nodes as well as to find (nearly) optimal immunization sets for a given choice of parameters and costs. The algorithm scales linearly with the size of the graph and it can be made efficient even on large networks. We compare its performance with topologically based heuristics, greedy methods, and simulated annealing.
\end{abstract}

\maketitle

\section{Introduction}\label{sec1}
One of the key questions of computational epidemiology is how best to distribute limited resources of treatment and vaccination so that they will be most effective in suppressing or reducing outbreaks
of disease. This problem is heightened by the entangled networks of interactions via which diseases can spread: in a large complex network, contact with a high-degree hub can see a virus
spread rapidly throughout the population even if the probability of transmission from an individual contact is low \cite{pastor-satorras_epidemic_2001,pastor-satorras_epidemic_2001-1}.
Early works on network immunization drew attention to the differences between random immunization and targeted immunization strategies \cite{pastor-satorras_immunization_2002,holme_attack_2002,cohen_efficient_2003,holme_efficient_2004}. A simple random immunization strategy can consist in fixing a fraction or a density of immunized nodes and averaging the outcome of the epidemic process over all possible realizations of the immunization set. On the contrary, targeted immunization strategies correlate the choice of immunized nodes with some topological feature, such as the degree or other centrality measures. This can be experimentally shown to have some positive effect in reducing the spread of diseases \cite{pastor-satorras_immunization_2002,holme_attack_2002,cohen_efficient_2003,holme_efficient_2004}. Most topologically-based algorithms for immunization follow an incremental procedure, in which the set of immunized nodes is initially empty then it is progressively filled adding one by one the nodes that are most relevant with respect to a particular topological metric. Despite the computational cost, recalculating the topological metric after each immunization step (i.e. after removing the immunized node from the graph) usually provides much better results than computing it only once on the original graph \cite{holme_attack_2002}. Further improvements were obtained by means of more complex immunization strategies, based on graph partitioning \cite{chen_finding_2008,hadidjojo_equal_2011} and on the optimization of the susceptible size \cite{schneider_suppressing_2011}.
Beside the heterogeneity of contacts, also clustering, community structure and modularity have a major impact on disease dynamics \cite{masuda_immunization_2009,salathe_dynamics_2010}, therefore the same immunization strategy can produce contrasting results on networks with different topological features. This is a consequence of the fact that topological heuristic methods neglect important features of the spreading rule and most common metrics used to measure their effectiveness, such as the largest connected component or the largest non-immune cluster size, are proxies that may not reflect the true susceptibility to an epidemic. These techniques also neglect the cost of vaccination, which may vary widely depending on the chosen target.

To overcome these limitations, several authors tried to quantify more explicitly the effects of immunization strategies on the outbreak dynamics \cite{borgs_how_2010,chung_distributing_2009,gourdin_optimization_2011,bauer_identifying_2012,preciado_optimal_2013,kashirin_heuristic_2013} and
 network immunization was mathematically formulated as a proper optimization problem, that can be proven to be NP-hard in a plethora of different variants \cite{kempe_maximizing_2003,giakkoupis_models_2005,aspnes_inoculation_2006,chen_better_2010,budak_limiting_2011,karkada_limiting_2011,nguyen_containment_2012,aditya_prakash_fractional_2013,preciado_optimal_2013}.
 Standard optimization techniques such as Monte-Carlo (MC) methods or integer/linear programming are computationally very expensive and may take a prohibitive amount of
time to reach reasonably good results even on relatively small networks. On the other hand, the {\em greedy} optimization strategies usually proposed are guaranteed to approximate the optimal result by a constant factor only in some fortunate case \cite{kempe_maximizing_2003,aditya_prakash_fractional_2013,preciado_optimal_2013}.

Recent progress in combinatorial optimization have shown that algorithms based on the message-passing principle, and developed using methods from the statistical physics of disordered systems, outperform in many cases both greedy algorithms and simulated annealing, even in complex optimization problem involving stochastic parameters \cite{altarelli_stochastic_2011,altarelli_stochastic_2011-1} and dynamical rules \cite{altarelli_large_2013,altarelli_optimizing_2012}. In many cases in which MC algorithms get trapped in local minima of the (free-)energy function, message-passing algorithms can find considerably better results. The remarkable performances of these algorithms are combined with considerably good computation time scaling properties. While on a wide variety of optimization problems the computational complexity of simulated annealing scales exponentially with the system size, message-passing algorithms typically require a time that scales roughly linearly with the number of messages (i.e. the number of edges).

In this paper we show that, under some approximations, network immunization can be written as a constrained optimization problem, in which the constraints are fixed-point equations for some local (node or edge) variables describing the average stationary state of the dynamics. These constraints and a suitably defined objective (energy) function are then used to derive a message-passing approach to the optimization problem, and to design efficient algorithms on large networks. We apply this method to find optimal immunization strategies for both susceptible-infected-recovered (SIR) and susceptible-infected-susceptible (SIS) models.

In Section \ref{sec2} we recall the main ideas and formulas of mean-field methods in epidemic models, that are usually used to estimate the average stationary properties of an epidemic outbreak. The optimal immunization problem is introduced in Section \ref{sec3}, opportunely defined in terms of mean-field quantities. Section \ref{sec4} is devoted to the definition of the message-passing approach and the derivation of the corresponding Belief-Propagation (BP) and Max-Sum (MS) equations.  In Section \ref{sec5}, we use BP to understand in detail the immunization properties on the prototypical case of random regular graphs.
The comparison with other optimization methods on more general graphs is discussed in Section \ref{sec6}.

\section{Mean-field methods in epidemic models}\label{sec2}
Over the years, a large number of stochastic epidemic models have been introduced, with the aim of addressing some specific features of different diseases \cite{anderson_infectious_1991,hethcote_mathematics_2000}. In the most simple model, the epidemic spreading induces in the nodes irreversible stochastic transitions from a {\em susceptible} state to an {\em infected} one. Infected individuals can   {\em recover} either returning to the susceptible state or becoming permanently resistant to the disease. 
One can then increase the complexity of the stochastic model introducing other intermediate states, or compartments, such as {\em exposure} and {\em latency}.
In the following, we discuss the most basic models of epidemic spreading, providing for each of them a set of approximated equations of mean-field type valid on very general graph structures. Their solution describes the statistical properties of the stationary state corresponding to a given set of initial conditions and external parameters. In addition, such equations allow to measure the level of infection once a configuration of initially immunized nodes is chosen.

\subsection{Irreversible Epidemic Processes}
The {\em susceptible-infected-recovered} (SIR) model was formulated by Kermack and McKendrick \cite{kermack_contribution_1927} to describe the irreversible propagation through a population of individuals of an infectious disease, such as measles, mumps, or cholera. The SIR stochastic dynamics is defined over a graph $G=(V,E)$, representing the contact network of a set $V$ of individuals. At any
given time step $t$ (e.g. a day), a node $i$ can be in one of three states: susceptible ($\cal S$), infected ($\cal I$), and recovered/removed ($\cal R$). The state of node $i$ at time $t$ is represented by a
 variable $x_i^t \in \{\cal{S,I,R}\}$. We assume that each node $i$ is initially infected with probability $q_i \in [0,1]$ (independent of the other nodes).  At each time step, an infected node $i$ can first spread the disease to each susceptible neighbor $j$ with given probability $T_{ij}\in (0,1]$, then recover with probability $r_i$. Once recovered, individuals do not get sick anymore (they are effectively removed from the graph). The probability that an infected node $i$ directly transmits the disease to $j$ before $i$ recovers is given by
\begin{eqnarray}
\nonumber p_{ij} & = & 1- \sum_{t=1}^{\infty} (1-T_{ij})^{t} r_i (1-r_i)^{t-1} \\
& =&  \frac{T_{ij}}{T_{ij}+(1-T_{ij})r_i}. \label{pij}
\end{eqnarray}
It is thus possible to construct a completely static representation of the process that maps the final state onto the outcome of a bond percolation process \cite{newman_spread_2002}. This relationship can be made mathematically clear as follows. Let us consider a tree-like graph and define $m_{ij}$ to be the probability that node $i$ is eventually infected when considering the graph obtained in the absence of the neighboring node $j$. Exploiting the factorization of probabilities on the sub-branches of the tree emerging from $i$, the quantity $m_{ij}$ satisfies the equation
\begin{equation}\label{sir-st0}
m_{ij} = q_i + (1- q_i)\left[1-\prod_{k\in \partial i\sm j}(1-p_{ki} m_{ki})\right]
\end{equation}
where $\partial i$ denotes the set of neighbors of $i$.
Since infected nodes eventually recover, in the final state nodes can only be either in the susceptible state or in the recovered one. From the knowledge of the conditional marginals $m_{ij}$, one can compute the probability $m_i$ that a node $i$ is eventually infected, i.e. the probability that $i$ is recovered in the final state,
\begin{equation}\label{sir-st1}
m_i = q_i + (1- q_i)\left[1-\prod_{k\in \partial i}(1-p_{ki} m_{ki})\right].
\end{equation}

\begin{figure}
\includegraphics[width=1\columnwidth]{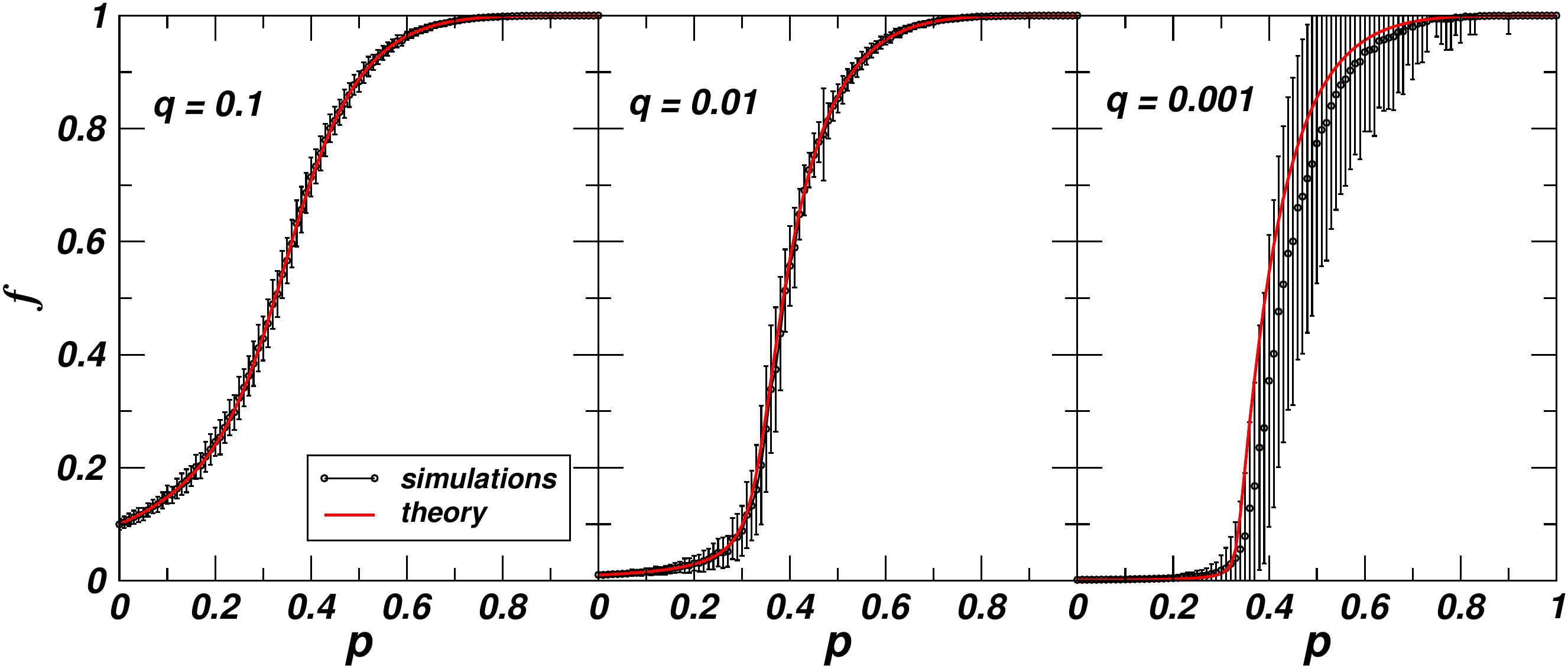}
\caption{Black circles represent the average density $f$ of nodes that have been infected by the final (infinite) time in the SIR dynamics on a random regular graph of $N=10^3$ nodes and degree $K=4$, as function of the transmission probability $p$ for $q=0.1,0.01,0.001$ (from left to right). The symmetric bars indicate the fluctuations around the average value computed on $10^4$ realizations of the stochastic process. The red full line is computed from the solution of \eqref{sir-st0}-\eqref{sir-st1}.
\label{fig-SIR-Ip}}
\end{figure}

Although \eqref{sir-st0}-\eqref{sir-st1} are exact only on trees, they have been successfully applied to study the SIR model also on general random graphs \cite{newman_spread_2002,dallasta_inhomogeneous_2005,karrer_message_2010}. A comparison between the solutions of these equations and the results of simulations of the SIR stochastic process is shown in Figure \ref{fig-SIR-Ip} for a random regular graph (RRG) of $N=10^3$ nodes and degree $K=4$. For simplicity we considered uniform self-infection probabilities $q_i = q$, $\forall i\in V$ and uniform transmission probabilities $p_{ij} = p$, $\forall (i,j)\in E$. In the SIR stochastic process, we defined a measure of  the ``outbreak'' size as the average fraction $f$ of nodes that have been infected during the epidemic spreading. Since all infected nodes eventually recover, this metric can be also defined as
\begin{equation}
f = \frac{1}{N}\sum_i \Pr\left[x_i^\infty=\cal R\right]
\end{equation}
where $\Pr\left[x_i^\infty=\cal R\right]$ is the probability that node $i$ is eventually recovered. In Fig.\ref{fig-SIR-Ip}, $f$ is plotted as function of the transmission probability $p$ for $q=0.1,0.01,0.001$. The results, obtained averaging over $10^4$ realization of the stochastic process, are reported as black circles, while the symmetric bars indicate the fluctuations around the average behavior. The average behavior can also be computed from the solutions of \eqref{sir-st0}-\eqref{sir-st1}, exploiting the fact that, in the tree-like approximation $\Pr\left[x_i^\infty=\cal R\right] \simeq m_i$. The results are reported as  a red full line in Fig.\ref{fig-SIR-Ip}.
The agreement between the mean-field theory represented by \eqref{sir-st0}-\eqref{sir-st1} and the simulations is very good for sufficiently large values of $q$, then it deteriorates for $q$ of the order of $1/N$ and large values of $p$. The reason for such a discrepancy is that Eqs.\eqref{sir-st0} are correct on tree-like structures, i.e. when the disease transmission events to one node coming from two neighbors are not correlated. The ``decorrelation'' assumption is not correct when the actual number of sources of spontaneous infection is very small. This is obvious in the case of a unique source of infection:  the contagion path has the same source, hence the infection of a node due to disease transmission from her neighbors is a highly correlated process, that is not well captured by \eqref{sir-st0}-\eqref{sir-st1}. More precisely the solution to Eqs. \eqref{sir-st0}-\eqref{sir-st1} gives an upper bound for the real probability to be infected \cite{karrer_message_2010}. In the limit of infinitely large networks, this approach is expected to provide a correct description of  the average final state of the system for any finite value of $q$ and $p$.

Equations \eqref{sir-st0}-\eqref{sir-st1} can be easily modified to include immunization of nodes. By considering a set of binary variable $s_i \in \{0,1\}$, in which $s_i =1$ if node $i$ is immune to the disease, we get $\forall i\in V$ and $\forall (i,j)\in E$
\begin{subequations}
\begin{equation}\label{sir-st2}
m_{ij} = (1-s_i) \left\{q_i + (1- q_i)\left[1-\prod_{k\in \partial i\sm j}(1- p_{ki}m_{ki})\right] \right\},
\end{equation}
\begin{equation}\label{sir-st3}
m_{i} = (1-s_i) \left\{q_i + (1- q_i)\left[1-\prod_{k\in \partial i}(1- p_{ki}m_{ki})\right]\right\}.
\end{equation}
\end{subequations}
Given a configuration $\Bs = \{s_1, \dots, s_N\}$ of immune nodes, that we call {\em immunization set}, the solution of \eqref{sir-st2}-\eqref{sir-st3} provides a measure of the corresponding epidemic outbreak. It is possible to show that for a given set of parameters $\{q_i\}$ and $\{p_{ij}\}$, the solution of \eqref{sir-st0}-\eqref{sir-st1} is unique, therefore each configuration of immune nodes $\Bs$ corresponds to a unique solution of the equations \eqref{sir-st2}-\eqref{sir-st3}. This property will be crucial for the validity of the optimization method developed in this work.

\subsection{Reversible Epidemic Processes}
The {\em susceptible-infected-susceptible} (SIS) model is the prototype of reversible models of epidemic spreading, in which after recovery a node is again susceptible of being infected \cite{anderson_infectious_1991,hethcote_mathematics_2000}. The state of node $i$ at time $t$ is now represented by a binary variable $x_i^t \in \{\cal{S,I}\}$. At each time step, an infected node $i$ can transmit the disease to each of its susceptible neighbors $j$ with probability $p_{ij}$, while it recovers with rate $r_i$ (becoming susceptible again). The stochastic process admits an absorbing state in which all nodes are susceptible and the disease has disappeared from the population. When the transmission probabilities are sufficiently large, an active stationary state also exists, that is metastable and attractive for the dynamical process. Although in any finite population a fluctuation will eventually bring the system into the absorbing state, the lifetime of the metastable endemic state scales with the size of the graph in such a way that an absorbing phase transition as function of the transmission probabilities is expected to occur in the thermodynamic limit \cite{,chakrabarti_epidemic_2008,castellano_thresholds_2010,mieghem_n-intertwined_2011,van_mieghem_viral_2012,mieghem_epidemic_2012}. The critical threshold usually depends on the parameters of the dynamical process as well as on the topological structure of the underlying interaction graph. A variant of the SIS model with spontaneous self-infection was recently introduced \cite{van_mieghem_epidemics_2012,cator_susceptible-infected-susceptible_2013} in order to simplify the numerical and mathematical analysis of the model.
The presence of spontaneous self-infection destroys the absorbing state, whereas the metastable state becomes the (unique) stationary state of the dynamics. On the other hand, for a given small self-infection probability, the dynamics shows a clear boundary between low infection region and a region of global spreading as function of the transmission probabilities. By scaling down self-infection, one can extrapolate information on the epidemic phase transition occurring for zero self-infection, avoiding the problems associated with the existence of an absorbing state.

The SIS model on a given graph with $N$ nodes is a Markov chain with $2^N$ states, whose stationary probability distribution cannot be explicitly computed for large systems. A simple mean-field approximation that turned out to provide a good qualitative and quantitive description of the stationary state of the SIS model is obtained replacing the exact probability distribution by a product measure over the nodes of the graph \cite{mieghem_n-intertwined_2011,van_mieghem_viral_2012,mieghem_epidemic_2012,li_susceptible-infected-susceptible_2012,guerra_annealed_2010,goltsev_localization_2012}. This factorization, also known as the {\em $N$-intertwined model}, leads to a set of ``quenched'' mean-field equations for the evolution of single-node infection probabilities $m_i$ with $i\in V$. Notice the difference between the SIR and SIS cases: while in the SIR model $m_i$ indicates the (mean-field) probability that node $i$ is eventually infected before the final state, in the SIS model it represents the (mean-field) probability that $i$ is infected in the stationary state of the dynamics. In the ``quenched'' mean-field approximation, the infection probability of node $i$ at time $t$ satisfies the equation
\begin{equation}
 m_i^{t+1} =  \left(1-r_i\right) m_i^t  + \left(1-m_i^t\right)\left\{q_i + (1-q_i)\left[1- \prod_{j\in\partial i} (1- p_{ji} m_j^t)\right] \right\}
\end{equation}
where $p_{ji}$ is the transmission probability from $j$ to $i$, $q_i$ is the spontaneous self-infection probability.
In the stationary state, the mean-field variables $\{m_i\}_{i\in V}$ are given by the solution of the fixed-point equations
\begin{equation}\label{sis-st}
m_i = (1-s_i)\frac{q_i+ (1-q_i)\left[1-\prod_{j\in\partial i} (1- p_{ji} m_j)\right]}{r_i + q_i+ (1-q_i) \left[1-\prod_{j\in\partial i} (1- p_{ji} m_j) \right]}
\end{equation}
where $s_i \in\{0,1\}$ says whether node $i$ is immune or not. As for the SIS model, it is possible to show that the mean-field quantity $m_i$ gives an upper bound for the real value of the infection probability of node $i$ in the stationary state \cite{mieghem_n-intertwined_2011,van_mieghem_viral_2012,mieghem_epidemic_2012}. The approximation can be improved considering second-order quantities, i.e. deriving closed equations for single-point marginals and pair-correlations, but the actual form of these equations is not unique and depends on the moment closure approximation adopted \cite{mieghem_n-intertwined_2011,van_mieghem_viral_2012,mieghem_epidemic_2012}.
Nevertheless, in most cases equations \eqref{sis-st} already provide a very good description of the stationary state of the SIS stochastic process.

\begin{figure}
\includegraphics[width=1\columnwidth]{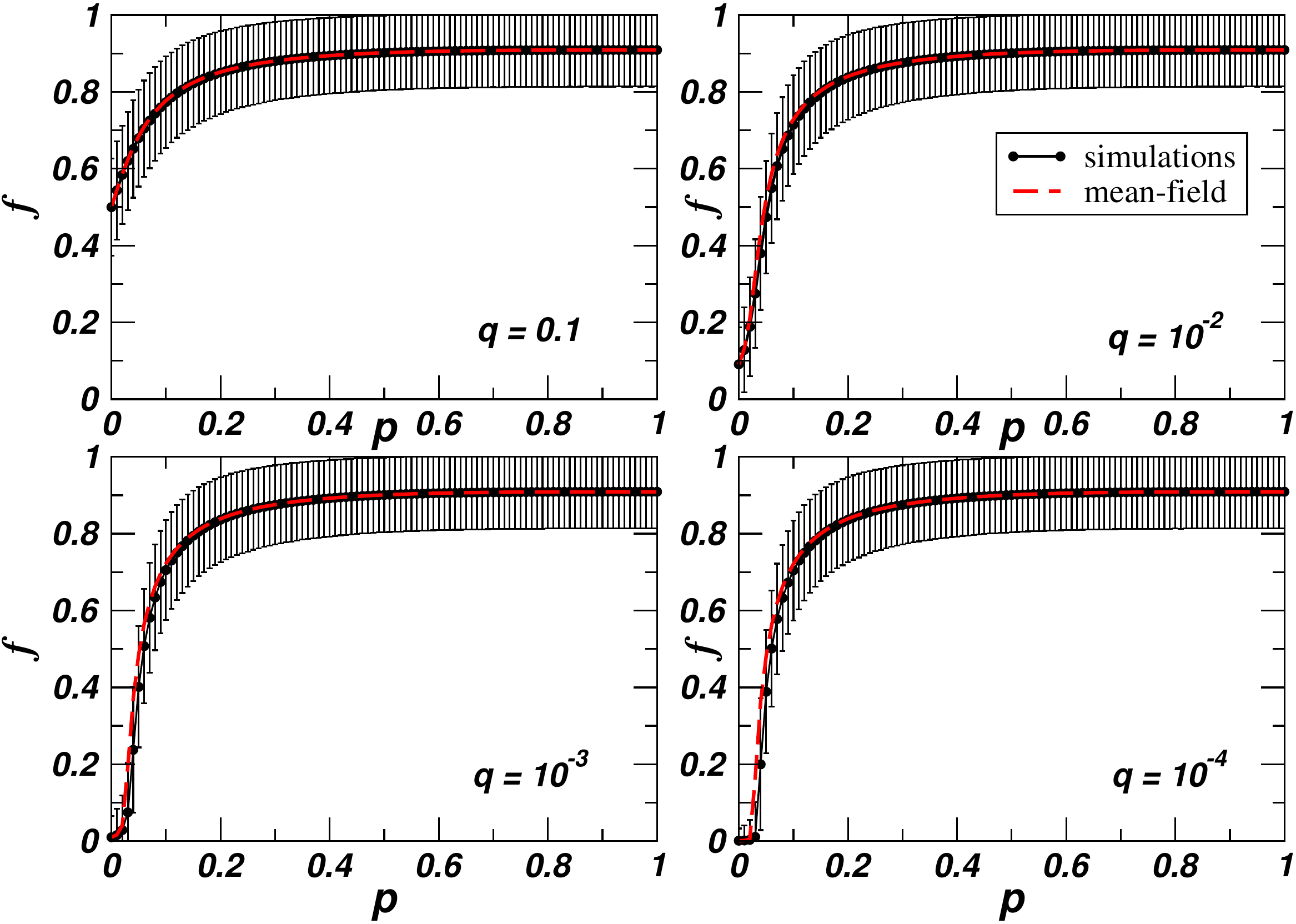}
\caption{Black circles represent the average density $f$ of infected nodes in the stationary state ($\tau = 10^4$ steps) of the SIS dynamics on a random regular graph of $N=10^3$ nodes and degree $K=4$, as function of the transmission probability $p$ for $q$ between $10^{-1}$ and $10^{-4}$. The symmetric bars indicate the fluctuations around the average value computed on $10^4$ realizations of the stochastic process. The red dashed line is computed from the solution of \eqref{sis-st}.
\label{fig-SIS-Ip}
}
\end{figure}

A measure of the outbreak size of the epidemics is given by the average fraction $f$ of infected nodes in the active stationary state. If the stationary state is infinitely long-lived, $f$ can be operatively defined as
\begin{equation}
f = \frac{1}{N}\sum_i\left\{\lim_{\tau \to \infty} \frac{1}{\tau}\sum_{t = 0}^{\tau} \1[x_i^t=\cal I]\right\}
\end{equation}
where $\1[\cdot]$ is equal to 1 when the argument is verified and 0 otherwise. In practice, simulations are performed for a finite time, that is chosen to be much longer than the time necessary to converge to the stationary state. Figure \ref{fig-SIS-Ip} displays the behavior of $f$ as function of the transmission probabilities $p$, for different values of the self-infection probability $q$, on  a random regular graph of $N=10^3$ nodes and degree $K=4$. The dashed line is the same quantity computed as $f \simeq \frac{1}{N} \sum_{i\in V} m_i$, i.e. approximating the probability that node $i$ is infected in the stationary state by the mean-field one $m_i$ obtained solving \eqref{sis-st}. The agreement between mean-field predictions and results of the simulations is very good in almost all regimes of the parameters.

\subsection{Continuous-Time Processes}

Both SIR and SIS epidemic models are often formulated in continuous time, where spontaneous self-infection probabilities and transmission probabilities are replaced by rates. While the time-dependent evolution of the dynamical processes can be quite different from that of their discrete-time counterpart, the stationary behavior at long-time can be similarly described in terms of percolation-like equations. In continuous time, transmission events can be modeled as independent Poisson processes, therefore the probability that, in a sufficiently small interval of time, a node is infected from the neighbors is proportional to the sum of the independent probabilities of the individual transmission events. Once $\{q_i\}_{i\in V}$ and $\{p_{ij}\}_{(i,j)\in E}$ are intended as probability rates, this reduces to replace the term $1- \prod_{j} (1- p_{ji} m_j)$ with $\sum_{j}p_{ji} m_j$  in both \eqref{sir-st2}-\eqref{sir-st3} and \eqref{sis-st}.

\section{The Optimal Immunization Problem}\label{sec3}

Preventing or eradicating diseases entails a trade-off between the costs of treating and hospitalizing infected individuals and the cost of distributing vaccines/drugs. When the contact network is known and these costs can be estimated, one can devise an optimization problem, in which the optimal immunization set is the configuration of immunized nodes that minimizes a properly defined energy function. In the SIR stochastic process we consider the following energy function
\begin{equation}
\mathscr{E}_{SIR}(\Bs)  =  \mu \sum_{i\in V} s_i c_i + \epsilon \sum_{i\in V} \ell_i \Pr\left[x_i^\infty=\cal R|\Bs\right]
\end{equation}
in which $c_i \in \mathbb{R}$ is the cost of immunization of node $i$, $\ell_i$ is a loss, i.e. the cost associated with the infection of node $i$, while $\Pr\left[x_i^\infty=\cal R|\Bs\right]$ is the probability that node $i$ eventually recovers (i.e. the probability that the node has been infected during the epidemic spreading) given the configuration $\Bs$ of immunized nodes. Estimating the probability $\Pr\left[x_i^\infty=\cal R|\Bs\right]$ from the simulations of the stochastic SIR process is very cumbersome, making the optimization problem practically unsolvable for sufficiently large graphs. In fact, it is known that finding the probability that a node becomes infected in the course of an epidemic in SIR and related models is an NP-hard problem \cite{shapiro_finding_2012}.
The energy function for the optimal immunization problem for the SIS model can be defined in a similar way to be
\begin{equation}
\mathscr{E}_{SIS}(\Bs)  = \mu  \sum_{i\in V} s_i c_i + \epsilon \sum_{i\in V} \ell_i \lim_{\tau \to \infty} \frac{1}{\tau}\sum_{t = 0}^{\tau} \1[x_i^t=\cal I].
\end{equation}
Likewise the SIR case, computing $\mathscr{E}_{SIS}(\Bs)$ from direct simulations of the stochastic dynamics is a very demanding task.

The problem of finding the configuration of seeds that minimizes an energy function associated with a given dynamical model has been investigated by several authors in the recent past.  Even for simple stochastic propagation models, such as the independent cascade model \cite{giakkoupis_models_2005}, several versions of this optimization problem have been introduced and shown to be NP-hard. To the best of our knowledge, the methods proposed to solve this class of problems are mostly based on greedy approaches, that scale reasonably well with the system size but give solutions that only in very peculiar cases provide good approximations of the real optima.

In this work we employ a simplified approach, based on the mean-field representation of the dynamical process.
We replace the energy functions $\mathscr{E}_{SIR}$ and $\mathscr{E}_{SIS}$ with a unique energy function defined as
\begin{equation}\label{energy}
\mathscr{E}(\Bs,\Bm)  = \mu\sum_{i\in V} s_i c_i + \epsilon \sum_{i\in V} m_i,
\end{equation}
in which $\forall i \in V$, $m_i$ is the solution of \eqref{sir-st2}-\eqref{sir-st3} and \eqref{sis-st} for the SIR and SIS models respectively.

This energy can now be computed easily for any configuration $\Bs$ of immunized nodes, solving the corresponding set of self-consistent equations for $\{m_i\}$. Once we know how to compute the energy function, the optimization problem can be studied with different methods. The simplest optimization method is a {\em greedy algorithm} that resembles the incremental procedure used in topologically-based heuristics. Starting from an empty set of immunized nodes, the greedy principle imposes to select the node whose immunization causes the largest drop in the energy function.
Once the node is included in the set of immunized nodes, a second node is chosen following the same greedy principle. One-by-one all nodes are added to the immunization set. Under this criterion, the best configuration of immunized nodes is the one that realizes the lowest energy. The greedy algorithm is fast but it presents several drawbacks: first, there is no guarantee that the best solution found is the optimal one; moreover, we cannot stop the greedy procedure before almost all nodes are added to the immunization set, because the energy function can display several local minima   during this process.
Another famous optimization method is {\em simulated annealing}, in which a MC algorithm is used to find the configuration $\Bs$ minimizing the energy $\mathscr{E}(\Bs)$. For a sufficiently slow annealing schedule, the algorithm is guaranteed of finding the minimum, i.e. the optimal immunization set. However, this can be computationally unfeasible in large graphs, since MC algorithms get trapped in local minima of the (free-)energy function and the convergence to the global one requires a time that can scale exponentially with the system size. In the next section, we will develop a message-passing approach to the optimal immunization problem.  Message-passing algorithms can find the global optimum or, in general, considerably better results than simulated annealing in a running time that scales only linearly with the number of messages (i.e. the number of edges). The implementation details of the greedy algorithm and simulated annealing are described in Appendix \ref{app-algo}.

\section{Belief-Propagation Approach to the Optimal Immunization Problem}\label{sec4}

Since each configuration of immunized nodes corresponds to a unique value of the energy function, we can define a probability function $Q(\Bs)$ by associating to each configuration $\Bs$ a Boltzmann weight  $e^{-\beta \mathscr{E}(\Bs,\Bm)}$ and tracing over the variables $\Bm$ with the constraint that they are a solution of the corresponding percolation-like equations.
It follows that the equations in \eqref{sir-st2} and in \eqref{sis-st} become a set of hard constraints for the auxiliary variables $\Bm$ that must be included in the optimization problem in addition to the energetic terms.
As for standard constraint-satisfaction problems over discrete variables (i.e. $\Bs$) defined on the vertices of a graph it is possible to apply the cavity method \cite{mezard_spin_1987,mezard_information_2009} and to develop efficient message-passing algorithms, such as BP and MS.

In the following we present the derivation of the BP and MS equations both for the SIR and SIS models, we discuss the approximation methods employed and the convolution technique used to  to solve them efficiently.

\subsection{Derivation for the SIR model}

\begin{figure}
\includegraphics[width=1\columnwidth]{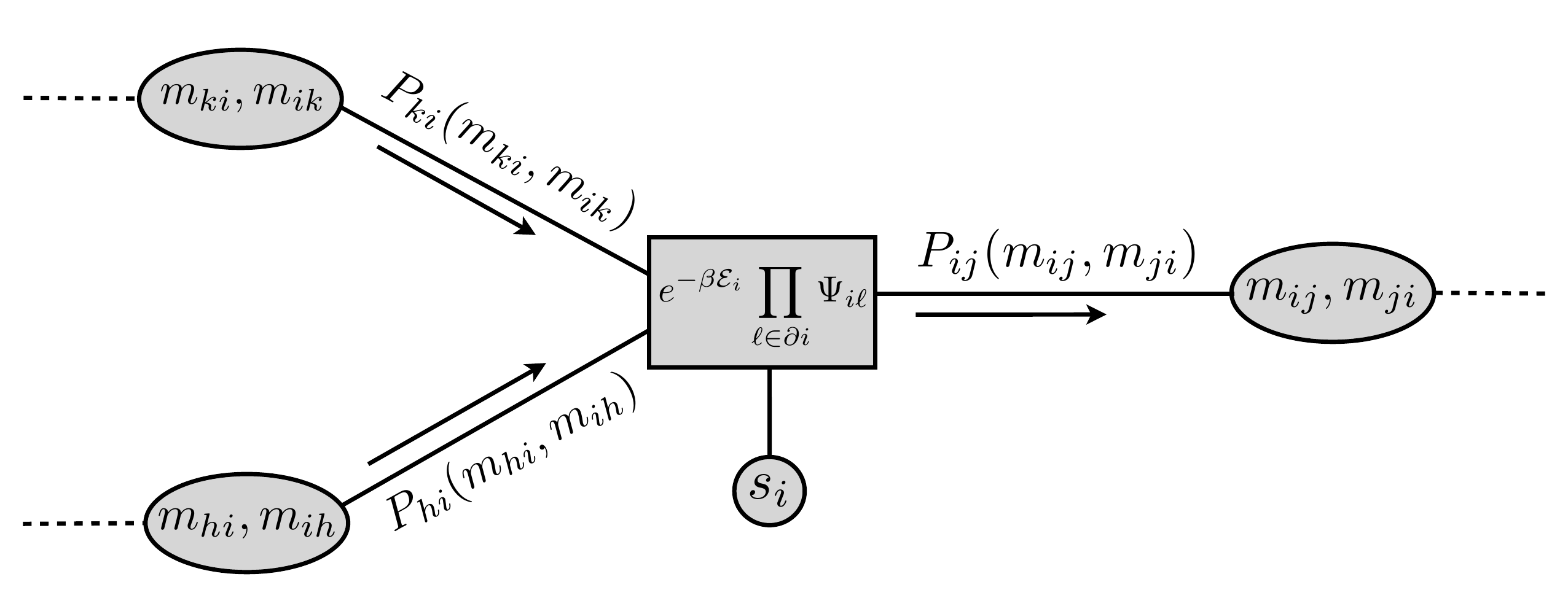}
\caption{Factor graph representation of the BP equations for the optimal immunization problem in the SIR model. Variable nodes are of two types: one type includes pairs of auxiliary variables $\{(m_{ij},m_{ji})\}$, the other includes the immunization variables $\{s_i\}$. There is a factor node for each node $i$ of the original graph,  including the energetic term $\mathscr{E}_i$ and all hard-constraints $\{\Psi_{i\ell}\}$ involving node $i$ as first label.
\label{fig-fgSIR}}
\end{figure}

In the SIR model, Equations~\eqref{sir-st2} define the values assumed by the auxiliary variables $\Bm$ for each configuration of immunized nodes $\Bs$. The relevant variables on which we must trace to compute the probability weight associated with configuration $\Bs$ are $\{ m_{ij}\}_{(i,j)\in E}$, with the condition that they satisfy  \eqref{sir-st2}.
Let us consider the partition function of the problem
\begin{equation}
Z = \sum_{\Bs} \int d\Bm e^{-\beta \mathscr E(\Bs,\Bm)}  \prod_{(i,j)\in E} \Psi_{ij}(s_i, m_{ij},\{m_{ki}\}_{k\in \dd i\sm j})
\end{equation}
where the integration over $d\Bm$ selects (for each $\Bs$) the unique set of values of $\Bm$ that satisfies the constraints
\begin{eqnarray}
\Psi_{ij}(s_i, m_{ij},\{m_k\}_{k\in \dd i\sm j}) =
 \delta\left(m_{ij}- (1-s_{i})\left\{1-(1-q_i)\prod_{k\in \dd i\sm j}(1-p_{ki}m_{ki})\right\}\right)
\end{eqnarray}
and where the energy is
\begin{eqnarray}\label{energySIR}
\nonumber \mathscr{E}(\Bs,\Bm) & = & \sum_i \mathscr{E}_i(s_i,m_i)\\
\nonumber & = & \mu\sum_i s_i c_i + \epsilon \sum_i \ell_i m_i\\
\nonumber & = &  \mu\sum_i s_i c_i + \epsilon \sum_i \ell_i(1-s_{i})\left\{q_i  + (1-q_i)\left[1-\prod_{k\in \dd i}(1-p_{ki}m_{ki})\right]\right\}.
\end{eqnarray}
The probability associated with a configuration of immunized nodes $\Bs$ is
\begin{equation}\label{probSIR}
Q(\Bs) = \frac{1}{Z}  \int d\Bm e^{-\beta \mathscr E(\Bs,\Bm)}  \prod_{(i,j)\in E} \Psi_{ij}(s_i, m_{ij},\{m_{ki}\}_{k\in \dd i\sm j}).
\end{equation}
In order to derive the BP equations, we consider an infinite tree, and we marginalize over all variables but $j$. In this way we obtain the probability $Q_j(s_j)$ for the variable $s_j$,
\begin{equation}\label{psSIR}
Q_j(s_j) \propto  \prod_{i\in \dd j} \int_{0}^{1}dm_{ij} \int_{0}^{1}dm_{ji} \ e^{-\beta \mathscr{E}_j(s_j,m_j)} \prod_{i\in\dd j} \Psi_{ji}(s_j, m_{ji},\{m_{kj}\}_{k\in \partial j\sm i}) \prod_{i\in\partial j} P_{ij}(m_{ij},m_{ji})
\end{equation}
in which the factorization over the neighbors $i$ of $j$ comes from the fact that in a tree each sub-branch is independent of the others, and the partial partition function of the sub-branch is represented by the ``cavity marginal" or BP message $P_{ij}(m_{ij},m_{ji})$. The latter denotes the joint probability for the auxiliary variables $m_{ij}$ and $m_{ji}$ in absence of the hard-constraints over node $j$ (i.e. $\prod_{i\in\dd j} \Psi_{ji}$) and of the energetic term $\mathscr{E}_j$; they satisfy the BP equations
\begin{equation}
\label{bpSIR}
P_{ij}(m_{ij}, m_{ji})  \propto  \sum_{s_i} \prod_{k\in \partial i\sm j} \int_{0}^{1} dm_{ki} \int_{0}^{1} dm_{ik} e^{-\beta\mathscr{E}_i(s_i, m_i)} \prod_{\ell\in\dd i}\Psi_{i\ell}(s_i,m_{i\ell},\{m_{\ell' i}\}_{\ell'\in \partial i\sm \ell})\prod_{k\in \partial i\sm j} P_{ki}(m_{ki} ,m_{ik}).
\end{equation}
Unlike most common cases, the BP message $P_{ij}$ depends on both auxiliary variables defined in the edge $(i,j)$. This happens because both of them enter in the definition of the energetic term (in fact, they are both contributing to $m_j$). On the contrary the set of hard constraints $\prod_{i\in\partial j} \Psi_{ji}$ does not depend on the variable $m_{ij}$. It follows that in cases in which the energetic term $\mathscr{E}$ does not depend on $m_j$ (e.g. for $\epsilon=0$ in \eqref{energySIR}), it can be shown that $P_{ij}$ only depends on $m_{ij}$.

From the BP marginals we can compute the single-site total probability marginal $P_i(m_i)$,
\begin{equation}\label{pmSIR}
P_i(m_i) = \sum_{s_i} \prod_{k\in \partial i} \int_{0}^{1}  dm_{ki} \int_{0}^{1}  dm_{ik} e^{-\beta\mathscr{E}_i(s_i, \{m_{ki}\})} \prod_{k\in \partial i} \Psi_{ik}(s_i,m_{ik},\{m_{k'i}\}_{k'\in \partial i\sm k}) \prod_{k\in \partial i}P_{ki}(m_{ki} ,m_{ik})
\end{equation}
that represents the probability distribution of the values assumed by the variable $m_i$ on that node. The update rule of the BP equations \eqref{bpSIR} can be represented graphically on a factor graph as shown in Fig.\ref{fig-fgSIR}.

The cavity message $P_{ij}(m_{ij},m_{ji})$ is a real function defined on the square $[0,1]\times [0,1]$. Since there is no information on the shape of this function, we proceed discretizing the interval $[0,1]$ in a number $N_B$ of bins, and solving numerically \eqref{bpSIR} assuming that  $P_{ij}(m_{ij},m_{ji})$ is defined on a two-dimensional mesh of $N_B\times N_B$ identical cells.
Because of discretization, the hard constraints in \eqref{bpSIR} are only approximately satisfied. In the following, the integrals over $\Bm$ variables will be replaced by sums whenever we consider discretized versions of the equations. With a naive discretization, it is not easy to keep under control the propagation of the error during the iteration of the BP update rule. We also considered a relaxed version of the constraints for which we can bound the error due to the discretization,  however we checked numerically that the two versions of the algorithm give practically the same results.

The computation time of the update rule in the BP equations as written in \eqref{bpSIR} scales as $N_B^{k_i-1}$, where $k_i$ is the degree of node $i$, making the trace over all configurations of the neighbors practically unfeasible on most graphs. The computational complexity of the BP update rule can be considerably reduced by exploiting the properties of convolutions of messages.
For an arbitrary set $D \subseteq \partial i\setminus j$, we define the convolution function
\begin{equation}
M_D(S,T) = \sum_{\substack{m_{ki}, k\in D\ \text{s.t.}\\ \\ S = \prod_{k\in D}[1-m_{ki}p_{ki}]}} \prod_{k\in D} P_{ki}\left(m_{ki}, q_i + (1-q_i)\left[1- \frac{T}{1-m_{ki}p_{ki}} \right]\right)
\end{equation}
where $T = \prod_{k\in \dd i} \left[1-m_{ki}p_{ki} \right]$ so that $0 \leq T \leq S \leq 1$. Then, for any disjoint sets $D_1$ and $D_2$, we have
\begin{equation}
M_{D_1\cup D_2}(S,T) = \sum_{\substack{S_1,S_2\ \text{s.t.}\\ S = S_1 S_2}} M_{D_1}(S_1,T) M_{D_2}(S_2,T).
\end{equation}
One can start from the empty set and add one by one the elements of $\partial i\sm j$, using the convolution rules.
Finally, the convolution function over the complete set $\partial i\sm j$ can be used to compute the out-coming message as
\begin{equation}\label{BPconvSIR}
P_{ij}(m_{ij}, m_{ji})  \propto \mathcal{S}_{ij}(m_{ij},m_{ji}) + \mathcal{N}_{ij}(m_{ij},m_{ji})
\end{equation}
where the two terms, defined as
\begin{subequations}
\begin{eqnarray}
\mathcal{S}_{ij}(m_{ij},m_{ji}) & = & e^{- \beta \mu c_i}\prod_{k \in \dd i\sm j} \sum_{m_{ki}} P_{ki}(m_{ki},0) \\
\mathcal{N}_{ij}(m_{ij},m_{ji}) & = & e^{-\beta\epsilon \left\{1 -(1-m_{ji}p_{ji})\left(1-m_{ij}\right)\right\}} M_{\partial i\sm j} \left(1-\frac{m_{ij}-q_i}{1-q_i}, (1-m_{ji}p_{ji})\left(1-\frac{m_{ij}-q_i}{1-q_i}\right)\right),
\end{eqnarray}
\end{subequations}
refer to node $i$ being immunized or not (the proportionality symbol as usual means that the message has to be properly normalized).
Using the convolution trick, the computational complexity of the update rule on a node of degree $k$ reduces to $O(k N_B^3)$. The factor $N_B^3$ comes from the computation of the trace over the auxiliary variables $\{m_{ki}\}$ by means of a convolution function: for each value of the $N_B$ values taken by $T$, we have to sum over all $N_B$ values taken by $S_1$ and by $S_2$.

Finally, in order to explore directly the optimal immunization assignments, we can define the MS messages
\[
\hat{P}_{ij}(m_{ij},m_{ji}) = \lim_{\beta\to\infty}\frac{1}{\beta}\log{P_{ij}(m_{ij},m_{ji})}.
\]
For an arbitrary set $D \subseteq \partial i\setminus j$, we define the convolution function
\begin{equation}
\hat{M}_D(S,T) = \max_{\substack{m_{ki}, k\in D\ \text{s.t.}\\ S = \prod_{k\in D}[1-m_{ki}p_{ki}]}} \sum_{k\in D} \hat{P}_{ki}\left(m_{ki}, q_i + (1-q_i)\left[1- \frac{T}{1-m_{ki}p_{ki}} \right]\right)
\end{equation}
where $T = \prod_{k\in \dd i} \left[1-m_{ki}p_{ki} \right]$ so that $0 \leq T \leq S \leq 1$. Then, for any disjoint sets $D_1$ and $D_2$, we have
\begin{equation}
\hat{M}_{D_1\cup D_2}(S,T) = \max_{\substack{S_1,S_2\ \text{s.t.}\\ S = S_1 S_2}} \left\{\hat{M}_{D_1}(S_1,T) + \hat{M}_{D_2}(S_2,T)\right\}.
\end{equation}
The MS equations (in their discretized and efficient version) read
\begin{equation}\label{MSconvSIR}
\hat{P}_{ij}(m_{ij}, m_{ji})  = \max\left\{\hat{\mathcal{S}}_{ij}(m_{ij},m_{ji}),\hat{\mathcal{N}}_{ij}(m_{ij},m_{ji})\right\} + C_{ij}
\end{equation}
where $C_{ij}$ is an (irrelevant) additive constant, and
\begin{subequations}
\begin{eqnarray}
\hat{\mathcal{S}}_{ij}(m_{ij},m_{ji}) & = & \mu c_i + \sum_{k \in \dd i\sm j} \max_{m_{ki}} \hat{P}_{ki}(m_{ki},0) \\
\hat{\mathcal{N}}_{ij}(m_{ij},m_{ji}) & = & \epsilon \left\{1 - (1-m_{ji}p_{ji})\left(1- m_{ij}\right)\right\} + \hat{M}_{\partial i\sm j} \left(1-\frac{m_{ij}-q_i}{1-q_i}, (1-m_{ji}p_{ji})\left(1-\frac{m_{ij}-q_i}{1-q_i}\right)\right).
\end{eqnarray}
\end{subequations}

\subsection{Derivation for the SIS model}
\begin{figure}[t]
\includegraphics[width=0.8\columnwidth]{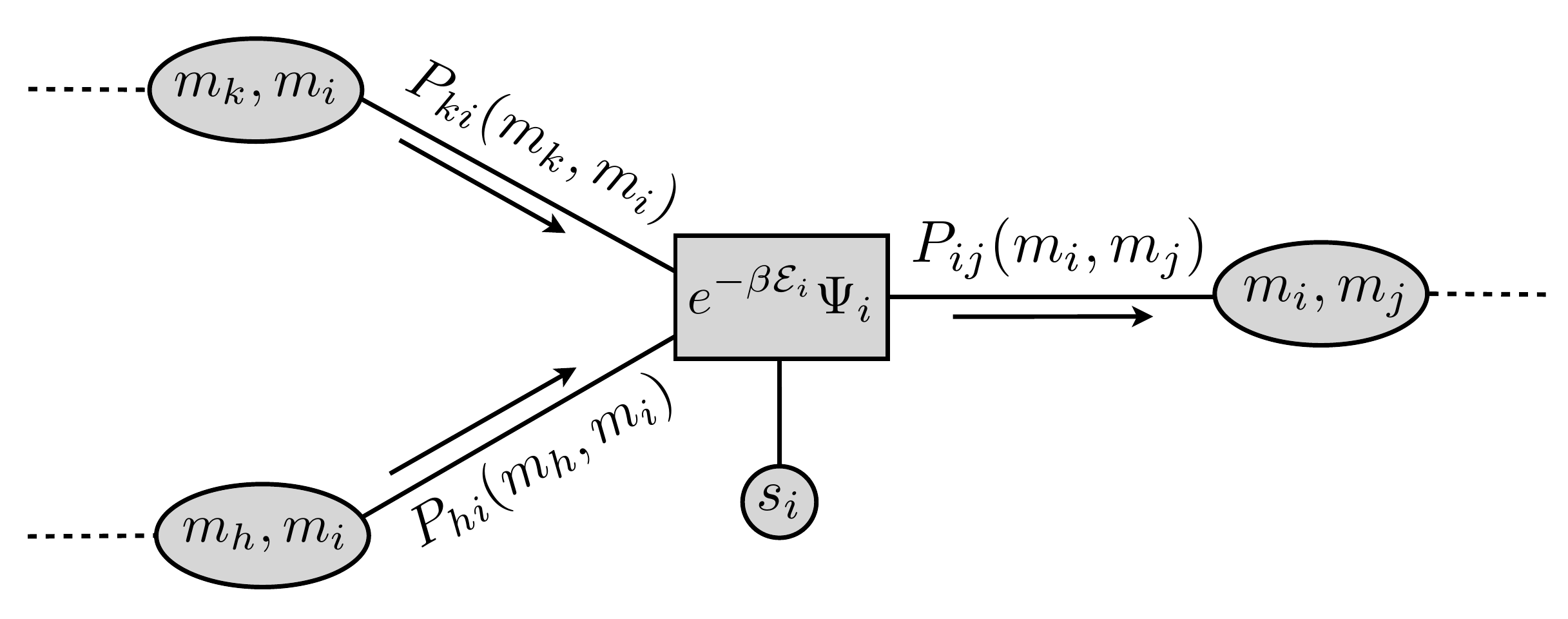}
\caption{Factor graph representation of the BP equations for the optimal immunization problem in the SIS model. Variable nodes are of two types: one type includes pairs of auxiliary variables $\{(m_{i},m_{j})\}$, the other includes the immunization variables $\{s_i\}$. There is a factor node for each node $i$ of the original graph, including the energetic term $\mathscr{E}_i$ and the hard-constraint $\Psi_{i}$.
\label{fig-fgSIS}}
\end{figure}

In the SIS model, the relevant auxiliary variables by means of which we can connect the configuration of immunized nodes with the corresponding energy value are the mean-field variables $\{m_i\}_{i\in V}$ satisfying \eqref{sis-st}. As done for the SIR model, we introduce Eqs.~\eqref{sis-st} as a set of hard constraints in a statistical mechanics model by means of  the partition function
\begin{equation}
Z = \sum_{\Bs} \int d\Bm e^{-\beta \mathscr E(\Bs,\Bm)}  \prod_{i\in V} \Psi_{i}(s_i, m_{i},\{m_{j}\}_{j\in \dd i}),
\end{equation}
where now we have
\begin{equation}
\Psi_{i}(s_i, m_{i},\{m_j\}_{j\in \dd i}) = \delta\left(m_{i}- (1-s_{i})\frac{q_i+(1-q_i)\left[1-\prod_{j\in \dd i}(1-p_{ji}m_{j})\right]}{r_i+q_i+(1-q_i)\left[1-\prod_{j\in \dd i}(1-p_{ji}m_{j})\right]}\right)
\end{equation}
and the usual energy term
\begin{eqnarray}
\mathscr{E}(\Bs,\Bm) & = & \sum_i \mathscr{E}_i(s_i,m_i) = \mu\sum_i s_i c_i + \epsilon \sum_i \ell_i m_i.
\end{eqnarray}
The probability associated to a configuration of immunized nodes $\Bs$ is
\begin{equation}\label{probSIS}
Q(\Bs) = \frac{1}{Z}  \int d\Bm e^{-\beta \mathscr E(\Bs,\Bm)}  \prod_{i\in V} \Psi_{i}(s_i, m_{i},\{m_{j}\}_{j\in \dd i}).
\end{equation}
We repeat the same derivation as before, assuming that the graph is an infinite tree. Marginalizing over all variables but $j$, we compute the probability $Q_j(s_j)$ for the variable $s_j$ as\begin{equation}\label{psSIS}
Q_j(s_j) \propto \int_{0}^{1}dm_j  \prod_{i\in \dd j} \int_{0}^{1}dm_{i} \ e^{-\beta \mathscr{E}_j(s_j,m_j)} \Psi_{j}(s_j, m_{j},\{m_{i}\}_{i\in \partial j}) \prod_{i\in\partial j} P_{ij}(m_{i},m_{j})
\end{equation}
where the BP messages $P_{ij}(m_i, m_j)$ satisfy the BP equations
\begin{equation}
\label{bpSIS}
P_{ij}(m_i,m_j) \propto  \sum_{s_i} \prod_{k\in \partial i\sm j} \int_{0}^{1} dm_k e^{-\beta\mathscr{E}_i(s_i, m_i)} \Psi_i(s_i,m_i,\{m_k\}_{k\in \partial i})\prod_{k\in \partial i\sm j} P_{ki}(m_k ,m_i)
\end{equation}
and it represents the joint probability that the mean-field variables on $i$ and $j$ assume values $m_i$ and $m_j$ in absence of the constraint $\Psi_j$ and of the energetic term $\mathscr{E}_j$.
Figure \ref{fig-fgSIS} displays a graphical representation of the BP equations \eqref{bpSIS} on the corresponding factor graph.

The single-site total probability marginal $P_i(m_i)$ is given by
\begin{equation}\label{pmSIS}
P_i(m_i) = \sum_{s_i} \prod_{k\in \partial i} \int_{0}^{1} dm_k e^{-\beta\mathscr{E}_i(s_i, m_i)} \Psi_i(s_i,m_i,\{m_k\}_{k\in \partial i})\prod_{k\in \partial i} P_{ki}(m_k ,m_i).
\end{equation}

As already described for the SIR model, the BP equations can be solved numerically using a discretized version of the messages, in which the $[0,1]$ interval is divided in $N_B$ bins. In this way, however, the number of operations required to compute the trace in \eqref{bpSIS} scales exponentially with the degree of the node, therefore we employ again the convolution method.
For an arbitrary set $D \subseteq \partial i\setminus j$, we define the quantity
\begin{equation}
M_D(S,m_i) = \sum_{\substack{\{m_{k}\}, k\in D\ \text{s.t.}\\ \\ S = \prod_{k\in D}[1-m_{k}p_{ki}]}} \prod_{k\in D} P_{ki}(m_{k}, m_{i}).
\end{equation}
Then, for any disjoint sets $D_1$ and $D_2$, we have
\begin{equation}
M_{D_1\cup D_2}(S,m_i) = \sum_{\substack{S_1,S_2\ \text{s.t.}\\ S = S_1 S_2}} M_{D_1}(S_1,m_i) M_{D_2}(S_2,m_i)
\end{equation}
and the BP equations become
\begin{eqnarray}
\label{bp}
\nonumber P_{ij}(m_i,m_j) & \propto & \sum_{s_i} \prod_{k\in \partial i\sm j} \int dm_k e^{-\beta\mathscr{E}_i(s_i, m_i)} \Psi_i(s_i,m_i,\{m_k\}_{k\in \partial i})\prod_{k\in \partial i\sm j} P_{ki}(m_k ,m_i)\\
\nonumber  & = &  \sum_{s_i} \sum_{S} M_{\dd i\sm j}(S,m_i) \delta\left(m_{i}- (1-s_{i})\frac{q_i+(1-q_i)\left[1-(1-p_{ji}m_{j})S\right]}{r_i+q_i+(1-q_i)\left[1-(1-p_{ji}m_{j})S\right]}\right) e^{-\beta\mathscr{E}_i(s_i, m_i)}  \\
& = & e^{-\beta\mu c_i}  + M_{\partial i\sm j} \left(\frac{1-m_i - r_i m_i}{(1-q_i)(1-m_i)(1-p_{ji}m_j)},m_i\right) e^{-\beta\epsilon\ell_i m_i}.
\end{eqnarray}

Again we can derive MS equations
\begin{equation}\label{MSconvSIS}
\hat{P}_{ij}(m_{i}, m_{j})  = \max\left\{\hat{\mathcal{S}}_{ij}(m_{i},m_{j}),\hat{\mathcal{N}}_{ij}(m_{i},m_{j})\right\} + C_{ij}
\end{equation}
where
\begin{subequations}
\begin{eqnarray}
\hat{\mathcal{S}}_{ij}(m_{i},m_{j}) & = & \mu c_i + \sum_{k \in \dd i\sm j} \max_{m_{k}} \hat{P}_{ki}(m_{k},0) \\
\hat{\mathcal{N}}_{ij}(m_{i},m_{j}) & = & \epsilon\ell_i m_i + \hat{M}_{\partial i\sm j} \left(\frac{1-m_i - r_i m_i}{(1-q_i)(1-m_i)(1-p_{ji}m_j)},m_i\right).
\end{eqnarray}
\end{subequations}
and $C_{ij}$ is an (irrelevant) additive constant.

\section{BP Results on Ensembles of Random Graphs}\label{sec5}

\begin{figure}[t]
\includegraphics[width=0.49\columnwidth]{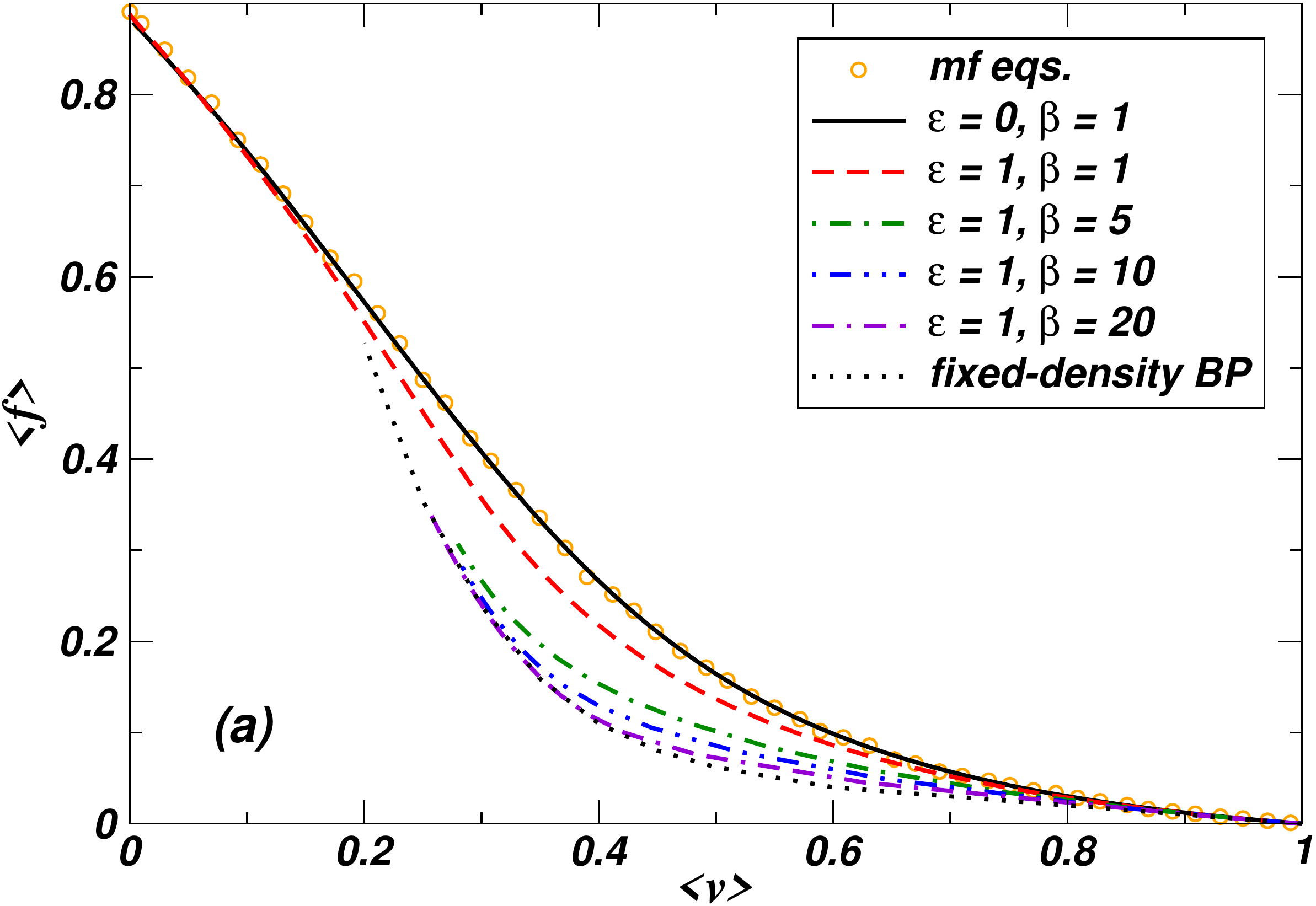}
\includegraphics[width=0.49\columnwidth]{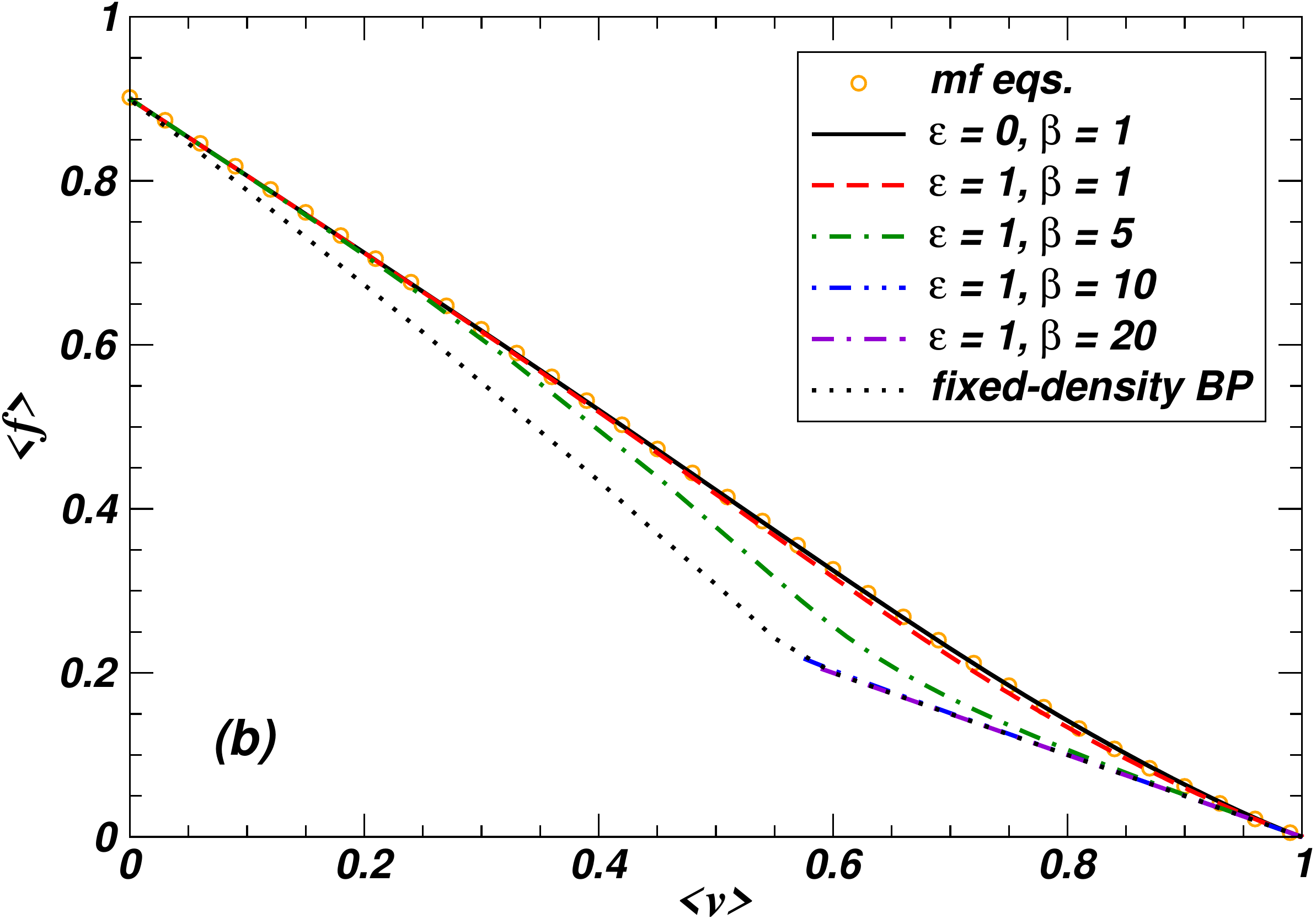}
\caption{Results for the analysis of the SIR and SIS models on random regular graphs of degree $K=4$, with uniform self-infection probability $q=0.1$ and uniform transmission probability $p=0.5$.
(a) For the SIR model, we plot the average density of nodes that got infected during the epidemic spreading $\langle f\rangle$ as function of the average density $\langle v\rangle$ of immunized nodes.
BP results on infinitely large graphs are reported for $\epsilon=0$ and $\beta=1$ (black full line) and for $\epsilon=1$ and $\beta=1$ (red dashed line), 5 (green dot-dashed line), 10 (blue double-dot-dashed line), and 20 (violet dot-double-dashed line). Results of sampling over Eqs.\eqref{sir-st2}-\eqref{sir-st3} (orange circles), corresponding to random immunization, are also displayed.
(b) For the SIS model, we  plot the average density $\langle f\rangle$ of infected nodes in the stationary state as function of the average density $\langle v\rangle$ of immunized nodes. BP results on infinitely large graphs are reported for $\epsilon=0$ and $\beta=1$ (black full line) and for $\epsilon=1$ and $\beta=1$ (red dashed line), 5 (green dot-dashed line), 10 (blue double-dot-dashed line), and 20 (violet dot-double-dashed line). Results of sampling over Eqs.\eqref{sis-st} (orange circles), corresponding to random immunization, are also displayed.
\label{fig-optBP}}
\end{figure}

On a general graph, the BP equations \eqref{bpSIR} and \eqref{bpSIS} are valid under the hypothesis of fast decay of correlations with the distance or replica symmetric (RS) assumption \cite{mezard_spin_1987,mezard_information_2009}. Under this assumption,  the statistical properties of the system are described by a unique Gibbs state (i.e. replica symmetry), and the BP equations admit a unique solution.
Random graphs are natural benchmark structures for evaluating the quality of the results obtained solving numerically the BP equations with histograms and the performances of the corresponding optimization method.  In order to isolate and study the effects of immunization on the statistical properties of epidemic spreading, we consider a completely homogeneous setup: a RRG with uniform values of both spontaneous self-infection and disease transmission along the edges, i.e. $q_i = q$, $\forall i\in V$ and $p_{ij} = p$, $\forall (i,j)\in E$. For the sake of simplicity, we also consider uniform loss parameters  and uniform immunization costs, i.e. $c_i = \ell_i=1$, $\forall i\in V$.
In the BP approach explained in Section \ref{sec4}, we can give a larger statistical weight to allocations of the immunized nodes that correspond to lower values of the energy. Increasing $\beta$ for $\epsilon> 0$ (see \eqref{energy}), the distribution $Q(\Bs)$ becomes biased towards immunization sets that generate a smaller expected number of infected nodes compared to random immunizations of the same density. In the limit of $\beta \to \infty$ the weight is concentrated on the minima of the energy function, i.e. on the optimal immunization sets.
In this framework an interesting global observable is the generalization of the quantity $f$ defined in Section \ref{sec3} when we perform an average over all possible immunization sets $\Bs$ with the corresponding weight $Q(\Bs)$. We call this quantity $\langle f\rangle$. We can exploit the definition of $f$ in terms of the variables $\Bm$, and use BP to obtain an estimate of $\langle f\rangle$,
\begin{equation}
 \langle f \rangle  =  \frac{1}{N} \sum_{i\in V} \langle m_i\rangle  =  \frac{1}{N} \sum_{i\in V}  \int_{0}^{1} dm_i P_i(m_i) m_i
\end{equation}
where $P_i(m_i)$ is given by \eqref{pmSIR} and \eqref{pmSIS} for the SIR and SIS models respectively.
The chemical potential $\mu$ can be used to control the average fraction of immunized nodes, denoted by $\langle v\rangle$, that is computed from the solution of the BP equations as
\begin{equation}
 \langle v \rangle  = \frac{1}{N} \sum_{i\in V} \langle s_i\rangle  =  \frac{1}{N} \sum_{i\in V}  \sum_{s_i=0,1} Q_i(s_i) s_i,
\end{equation}
and for $Q_i(s_i)$ we use \eqref{psSIR} and \eqref{psSIS} for the SIR and SIS models respectively.  It is thus possible to compute $\langle f\rangle$ as function of $\langle v\rangle$ for a fixed choice of the other parameters. We present here results obtained for infinitely large regular random graphs, obtained using the BP equations in the single-link approximation, i.e. when we solve self-consistently the BP equations assuming all nodes to have essentially the same statistical properties. For both SIR and SIS models, the results of the BP equations in the single-link approximation are shown in Fig.\ref{fig-optBP} with the choice of parameters $q=0.1$ and $p=0.5$ and different values of $\epsilon$ and $\beta$. In the case of random immunization ($\epsilon = 0$), the results obtained using BP on infinitely large RRG are compared with the average behavior observed by sampling the solutions of \eqref{sir-st2}-\eqref{sir-st3} (and \eqref{sis-st} respectively) and by simulating the SIR (and SIS) stochastic process on finite RRG of $N=10^3$ nodes. The latter are obtained sampling over $10^3$ configurations of immunized nodes for each value of $\langle v\rangle$. The agreement is very good for both models.

Increasing $\beta$ for $\epsilon=1$, the solutions of the BP equations show a monotonic decrease in the density of infected nodes.
The reduction of the infection level is particularly visible for intermediate values of $\langle v\rangle$, while it becomes almost negligible at small and large density of immunized nodes.
The fact that the density $\langle v\rangle$ is not directly fixed (as for micro-canonical systems), but it is implicitly varied as an effect of tuning the chemical potential $\mu$ and then evaluated from the outcome of the BP equations, is the cause of an undesirable issue at large values of $\beta$. It happens that, for large $\beta$, the free-energy of the statistical mechanics problem is not convex over the whole interval $[0,1]$ of values assumed by $\langle v\rangle$. The Legendre transform, implicitly used in the cavity method~\cite{mezard_information_2009}, spontaneously selects the convex envelope of the free-energy, limiting the interval of the possible values assumed by the density of immunized nodes. This is visible  in Fig.\ref{fig-optBP}, as increasing $\beta$ for $\epsilon=1$ the curves get interrupted at some (non-zero) value of $\langle v\rangle$. This means that, for that choice of the parameters, smaller non-zero immunization sets are, in energetic terms, less convenient than no immunization at all (i.e. of the point at $\langle v\rangle =0$). To eliminate this phenomenon, one can fix the value of $\langle v\rangle$ introducing an adaptive external field that is self-consistently adjusted during the iterations of the BP equations \cite{krzakala_elusive_2010}. Using this technique we were able to explore all values of the density of immunized nodes, although the procedure is not guarantee to converge at all values of $\beta$. More precisely, we computed the set of points indicated with black dots in Fig.\ref{fig-optBP}, that correspond to the best results (lowest value of $\langle f\rangle$) obtained varying $\beta$ at fixed values of $\langle v \rangle$. We expect these results to be very close to the optimal ones for a large graph. 

\begin{figure}
\includegraphics[width=0.49\columnwidth]{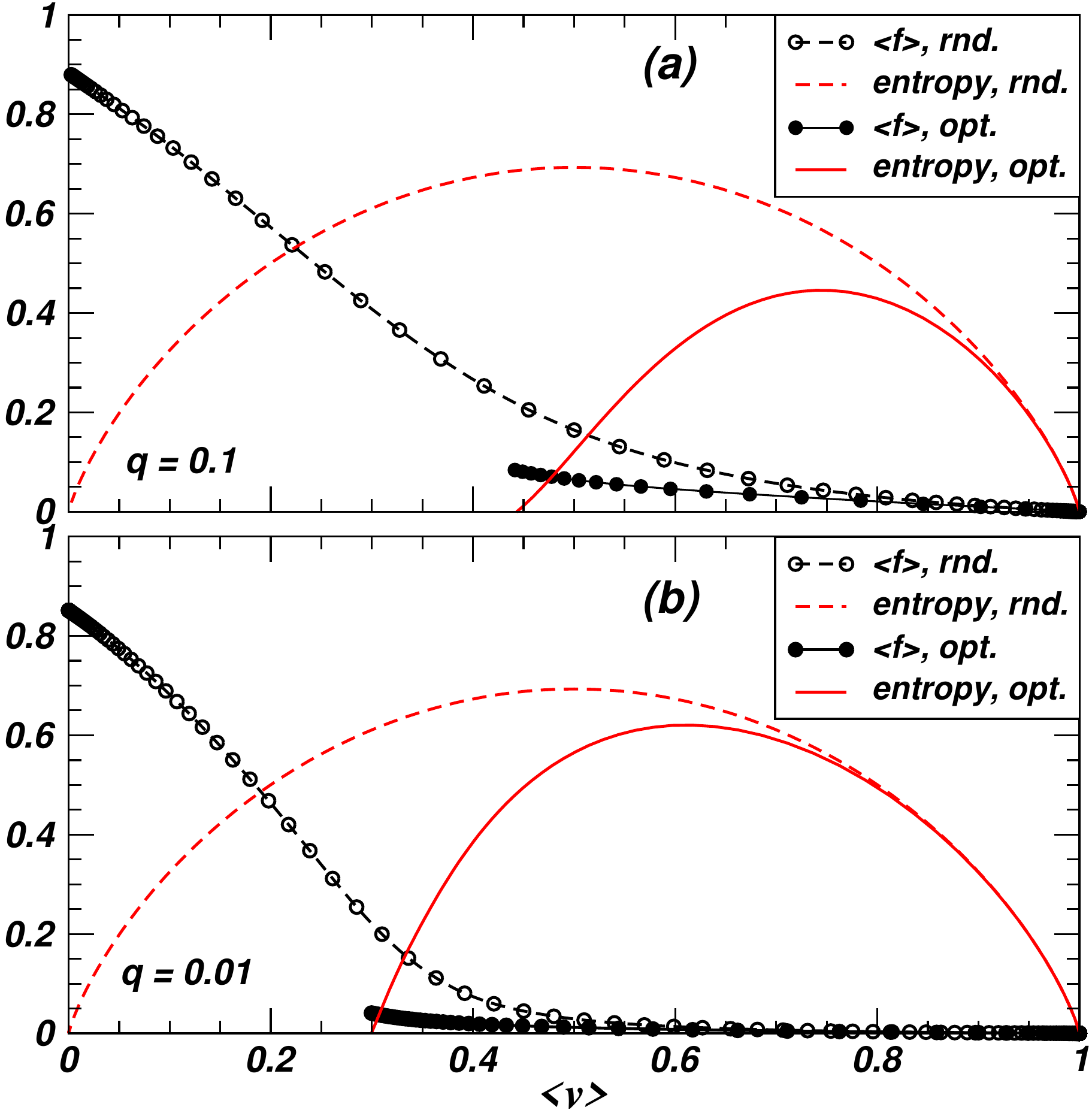}
\includegraphics[width=0.49\columnwidth]{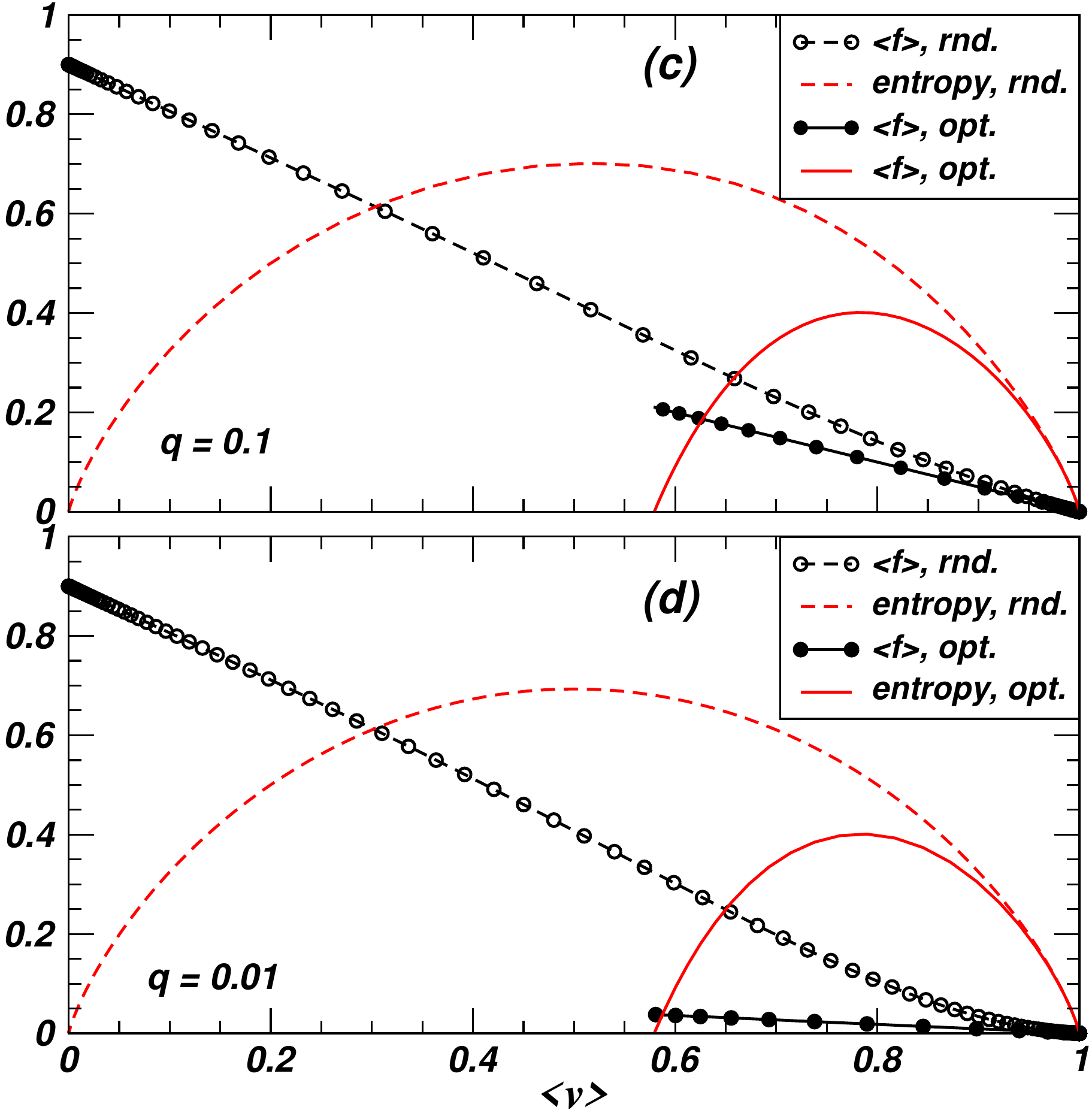}
\caption{Results for the analysis of the SIR and SIS models on random regular graphs of degree $K=4$, with uniform self-infection probability $q=0.1,0.01$ and uniform transmission probability $p=0.5$. For the SIR model (a-b), we plot the average density of nodes that got infected during the epidemic spreading $\langle f\rangle$ as function of the average density $\langle v\rangle$ of immunized nodes, both in the random case (open circles) and under optimization (full circles). For the SIS model (c-d), we plot the average density $\langle f\rangle$ of infected nodes in the stationary state as function of the average density $\langle v\rangle$ of immunized nodes, both in the random case (open circles) and under optimization (full circles).
\label{fig-optBP2}}
\end{figure}

Upon decreasing the spontaneous self-infection probability $q$, the effect of the optimization becomes more pronounced. This is shown in Fig.\ref{fig-optBP2}, where the results of optimization are compared with random immunization both for $q=0.1$ and $q=0.01$.
In the same Figure, we also plot the entropy $\mathcal{S}$ of the immunization sets as function of the density $\langle v \rangle$ (computed from the solution of the BP equations as in \cite{mezard_information_2009}). The red dashed lines, referring to random immunization, is the standard entropy curve given by $\mathcal{S}(\langle v\rangle) = -\langle v\rangle \log{\langle v\rangle} - (1-\langle v\rangle)\log{(1-\langle v\rangle)}$ and derived from a binomial distribution of $\langle v\rangle N$ immunized nodes among the $N$ possible ones. Red full lines instead represent the entropy curves under strong optimization, i.e. corresponding to low energy states. In this case, the entropy  becomes negative for sufficiently small values of $\langle v\rangle$, meaning that no configurations of immunized nodes can be found in the Gibbs state defined at that temperature (this is related to the non-convexity argument explained before). In other words, in that region it is (energetically) more convenient to let the disease spread naturally (with no immune) than performing immunization.

\begin{figure}[t]
\includegraphics[width=0.7\columnwidth]{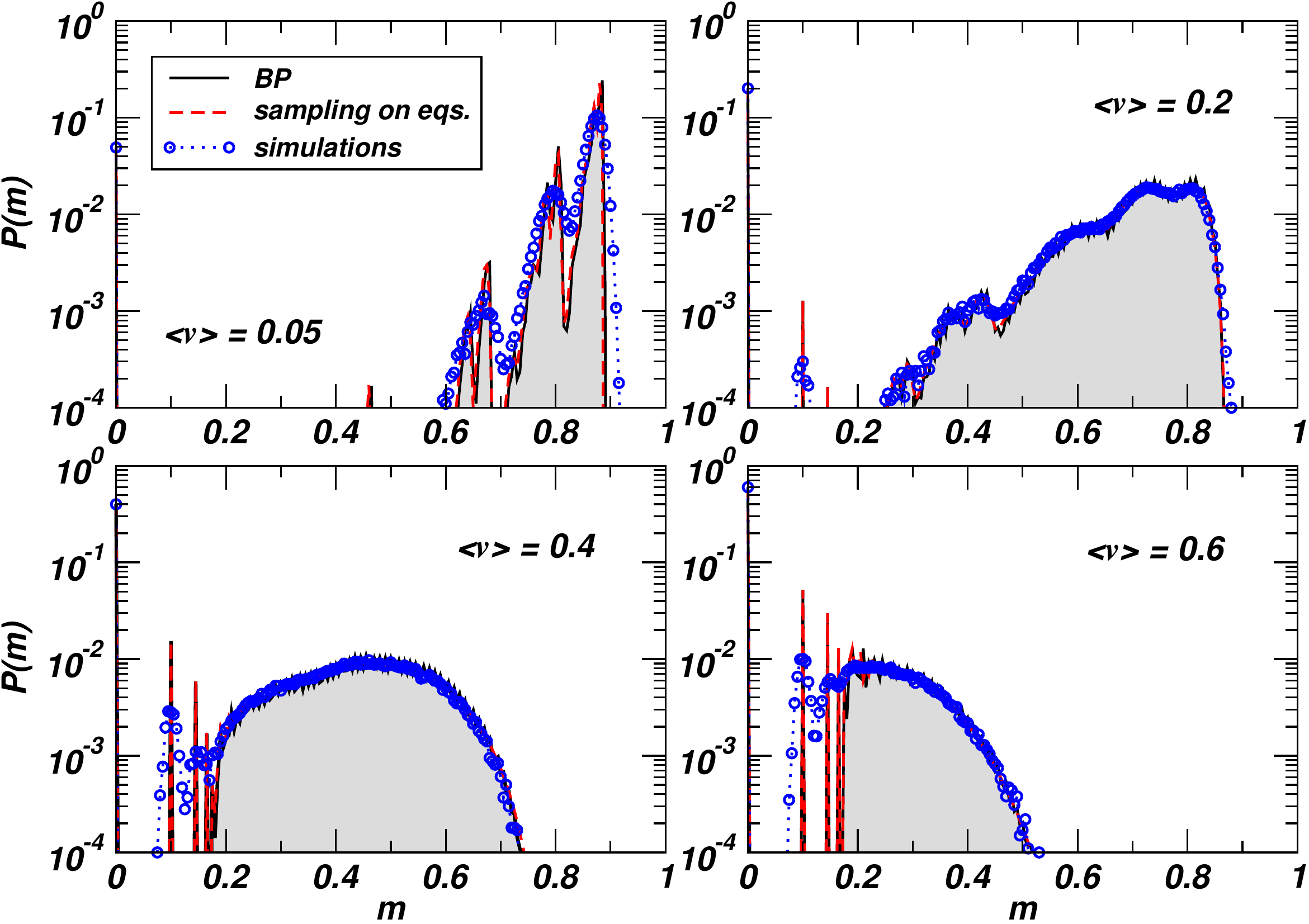}
\caption{Distribution $P(m)$ of the probability $m$ that a node got infected during the SIR process on random regular graphs of degree $K=4$ for $\langle v \rangle = 0.05,0.2,0.4, 0.6$. Results are computed using BP by means of \eqref{pmSIR} (black full line) on infinite networks, by sampling solutions of \eqref{sir-st2}-\eqref{sir-st3} (red dashed lines) and by means of simulations of the stochastic dynamics, both on a finite network of $N=10^3$ nodes and with a sample of $10^4$ immunization sets.
\label{fig-PmSIR}}
\end{figure}

With the BP analysis we can go beyond the study of the average macroscopic behavior and focus on more informative quantities, such as the distribution $P_i(m_i)$. In the absence of immunization, $P_i(m_i)$ is a delta function because the equations \eqref{sir-st2} (and \eqref{sis-st}) have a unique solution. After averaging over a set of configurations of immunized nodes, each node $i$ is represented by a distribution $P_i(m_i)$ instead of a single value of infection probability $m_i$. In a completely homogeneous setup (i.e. random regular graph and uniform parameters) all nodes have the same statistical properties, therefore we can ignore the node label in the distribution and assume $P_i(m_i) = P(m)$, $\forall i\in V$.
When the set of immunized nodes is drawn from a uniform distribution (i.e. $\epsilon = 0$), we can directly compare the shape of $P(m)$ obtained solving the BP equations with that obtained either by explicitly simulating the stochastic process or by repeatedly solving the mean-field equations with a large sample of configurations of immunized nodes.
Examples of the results obtained for the SIR model on RRG of degree $K=4$, $q=0.1$ and $p=0.5$ are displayed in Figure \ref{fig-PmSIR}. The shape of the distribution $P(m)$ varies considerably for different  values of the average immunization density $\langle v\rangle$. The comparison between the results for $\epsilon=0$ and the empirical distributions obtained by sampling solutions of \eqref{sir-st2}-\eqref{sir-st3} and by means of simulations of the stochastic dynamics is very favorable. The agreement between BP and  sampling results seems not to be affected by the discretization of the $[0,1]$ interval employed in order to solve the BP equations (here $N_B=200$). Also the agreement with the results of simulations is very good, and it improves increasing the number of configurations of immunized nodes employed in the simulations (here we performed an average over $10^4$ realizations at fixed density).

The distribution is always very heterogeneous and the average value (reported in Fig.\ref{fig-optBP}) is not at all representative of the behavior of the system. In general, $P(m)$ is characterized by a series of isolated peaks at low values of $m$ and by a continuous distribution in the bulk. The origin of the delta peaks is strictly related with the local structure of the graph around a node after immunization. For instance, the first peak for $m>0$ corresponds to isolated nodes, i.e. nodes that are completely surrounded by immunized ones and thus disconnected from the rest of the graph. Such nodes are infected with probability $q$, that is exactly the position of the first delta peak. A second peak occurs at $m=0.145$, corresponding to the probability of a node being part of an isolated infected dimer. For infected trimers (chain of length three), the central node corresponds to $m=0.187$, while the external nodes have $m=0.165$. These peaks are visible in Figure \ref{fig-PmSIR}. Then one can continue identifying other local clusters, such as star-like structures and small chains, each one corresponding to an isolated peak in the distribution. The continuous bulk of the distribution should be instead identified with a superposition of values due to large-scale clusters of infected nodes.

\begin{figure}
\includegraphics[width=0.7\columnwidth]{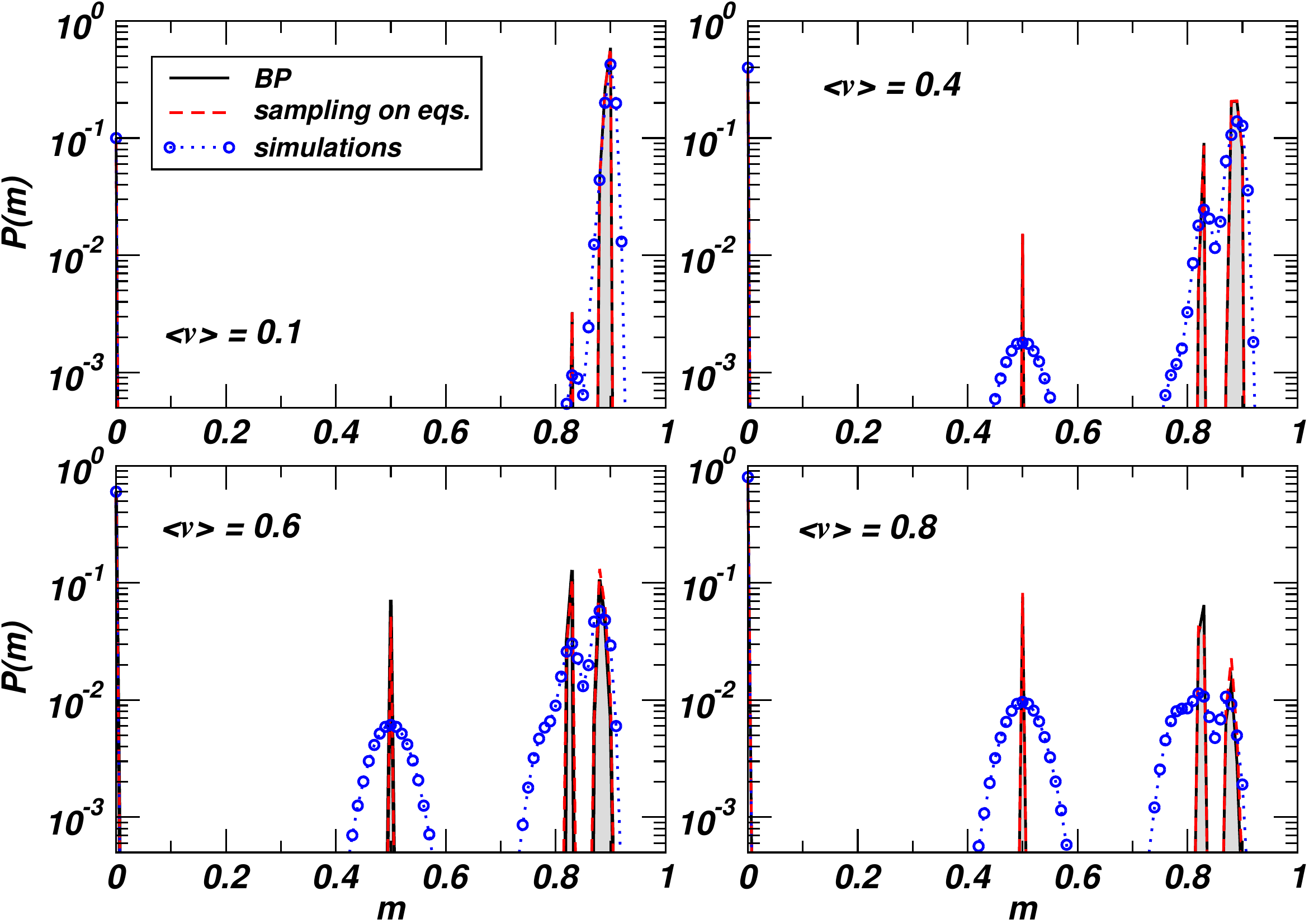}
\caption{Distribution $P(m)$ of the probability $m$ that a node is infected in the stationary state of the SIS process on random regular graphs of degree $K=4$ for $\langle v \rangle = 0.1,0.4, 0.6,0.8$. Results are computed using BP by means of \eqref{pmSIS} (black full line) on infinite networks, by sampling solutions of \eqref{sis-st} (red dashed lines) and by means of simulations of the stochastic dynamics, both on a finite network of $N=10^3$ nodes and with a sample of $10^4$ immunization sets.
\label{fig-PmSIS}}
\end{figure}

In Figure \ref{fig-PmSIS} we reported a similar plot for the SIS model. The structure of $P(m)$ is generally different, with few isolated peaks in which the weight of the distribution concentrates. This effect could be due to the naive mean-field approximation applied at the level of the equations used as hard-constraints, that is less accurate than that applied in the SIR model.

\begin{figure}
\includegraphics[width=0.49\columnwidth]{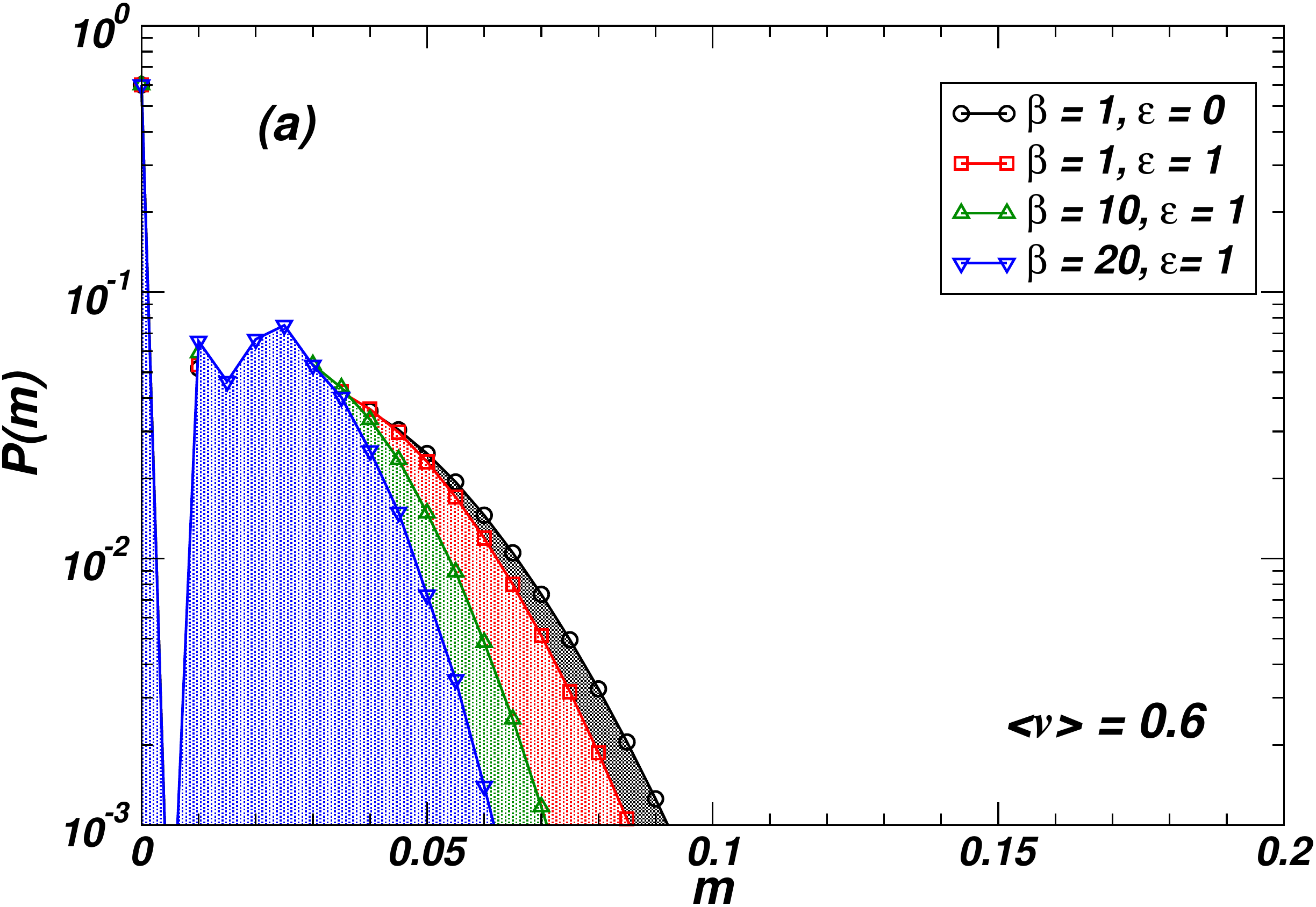}
\includegraphics[width=0.49\columnwidth]{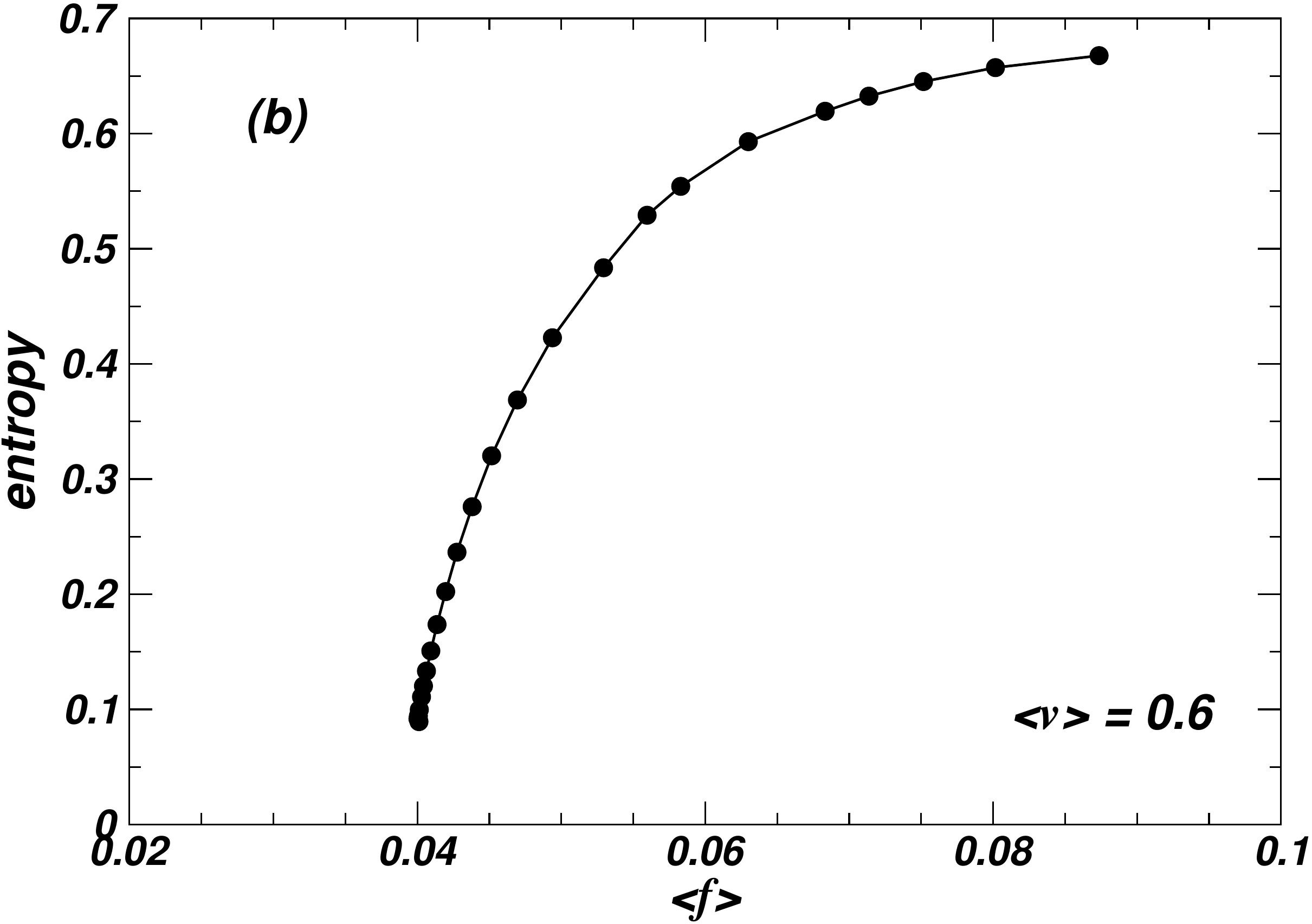}
\caption{(a) Distribution $P(m)$ of the probability $m$ that a node got infected during the SIR process on random regular graphs of degree $K=4$, $q=0.1$, $p=0.5$, $\langle v\rangle= 0.6$ and different values of $\beta$ and $\epsilon$. (b) Entropy of immunization sets of density $\langle v\rangle=0.6$ as function of $\langle f\rangle$ (obtained in creasing $\beta$ for $\epsilon=1$).
\label{fig-optPE-SIR}}
\end{figure}

\begin{figure}
\includegraphics[width=0.49\columnwidth]{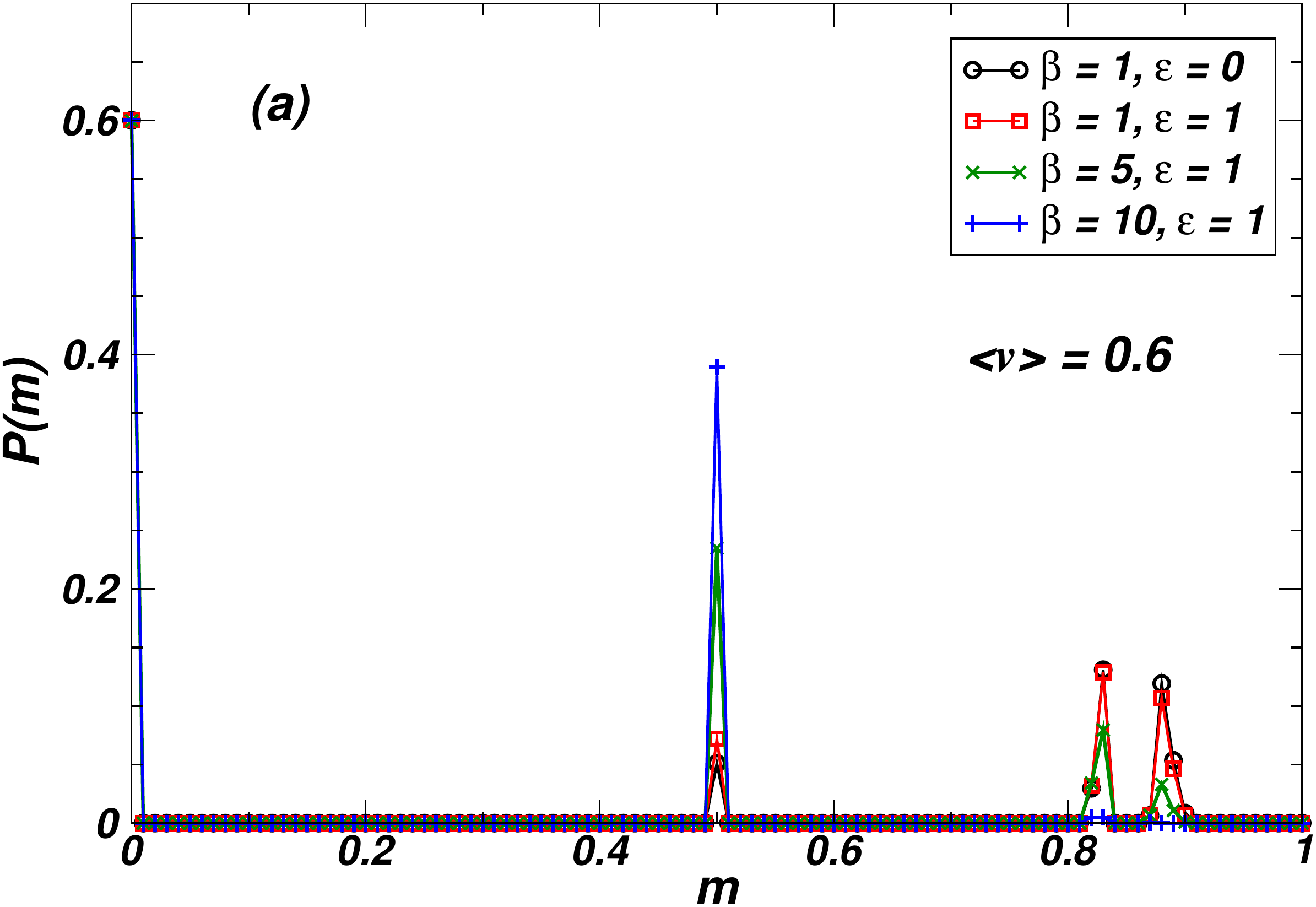}
\includegraphics[width=0.49\columnwidth]{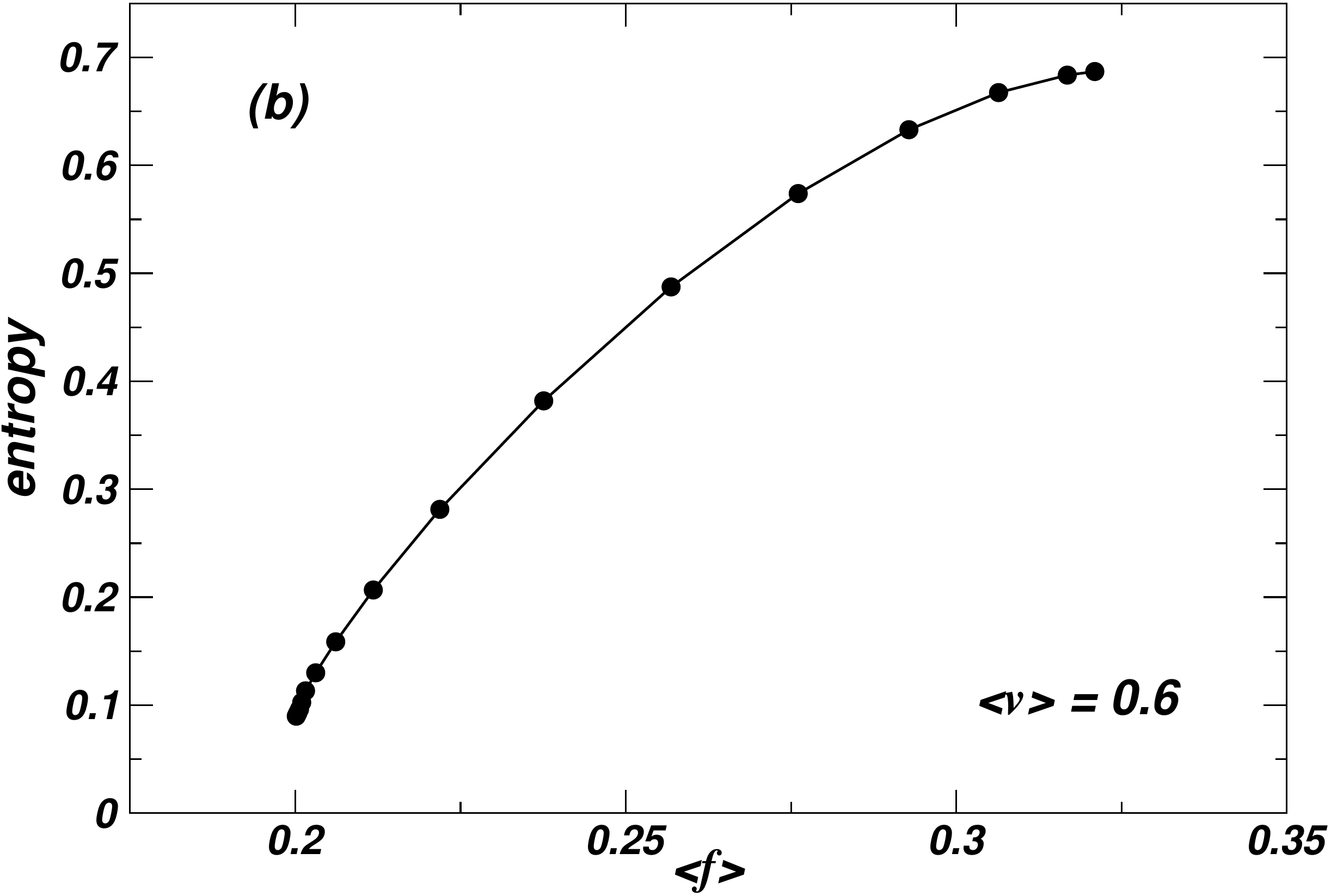}
\caption{(a) Distribution $P(m)$ of the probability $m$ that a node is infected in the stationary state of the SIS process on random regular graphs of degree $K=4$, $q=0.1$, $p=0.5$, $\langle v\rangle= 0.6$ and different values of $\beta$ and $\epsilon$. (b) Entropy of immunization sets of density $\langle v\rangle=0.6$ as function of $\langle f\rangle$ (obtained in creasing $\beta$ for $\epsilon=1$).
\label{fig-optPE-SIS}}
\end{figure}

There is a remarkable difference in the shape of $P(m)$ when a non-uniform weight is associated with immunization sets that generate different levels of infection. In the BP formalism, this is done by increasing $\epsilon$ from zero. Figure \ref{fig-optPE-SIR}a displays the distribution $P(m)$ for the SIR model at a fixed value of $\langle v\rangle = 0.6$, $\epsilon=1$ and different values of $\beta$. Increasing $\beta$ the distribution concentrates on a narrower interval of values of $m$. The information in these plots is important because it can be used to compute, for a given node in a graph, the probability that such node is infected for a given immunization strategy.
One can also ask how many immunization sets exist at a given density $\langle v \rangle$ that generate an average infection level of $\langle f \rangle$. For a RRG with degree $K=4$ and $\langle v \rangle = 0.6$, this is shown in Fig.\ref{fig-optPE-SIR}b, where we plot the entropy of immunization sets of fixed density as function of $\langle f\rangle$. These results are obtained from the solutions of the BP equations increasing $\beta$ from 1 (for $\epsilon=1$). In Figure \ref{fig-optPE-SIS} we report analogue plots for the SIS model with $\langle v \rangle = 0.6$. For a random allocation of immunized nodes (black circles in Fig.\ref{fig-optPE-SIS}a), $P(m)$ exhibits (in addition to a delta in zero) a series of narrow peaks at positive $m$. Increasing $\beta$, those at larger $m$ slowly disappear leaving only a delta peak at $m\approx 0.5$. Fig.\ref{fig-optPE-SIS}b shows the behavior of the entropy curve at $\langle v \rangle = 0.6$ for the SIS model. In both models, the entropy at $\langle v \rangle = 0.6$ does not vanish continuously when the minimum value of infection $\langle f\rangle$ is reached, but it remains considerably large. This is in accord with the behavior of the entropy curves displayed in Fig.\ref{fig-optBP2}. A continuously vanishing behavior can be observed instead if we select a density value that falls in the region in which, at large $\beta$, the entropy curve becomes negative (see later discussion about MS results). 

For $q=0.1$, the BP results have been obtained using a discretization of the interval $[0,1]$ in $N_B = 200$ bins. In general, the smallest possible non-zero value assumed by $m$ is $m=q$, therefore in order to have a sufficiently good resolution, one should always use $N_B > 1/q$. As the time complexity of the BP and MS algorithms scales as $N_B^3$, working with histograms when $q$ is small is a problem. A possible solution is that of adopting a mixed representation, such as
\begin{equation}
P(m) = \sum_{\ell=1}^{L} a_{\ell} \delta(m-m_{\ell}) + \sum_{\ell'=1}^{N_B-1} {b}_{\ell'}\1\left[m_{\ell'}< m \leq m_{\ell'+1}\right].
\end{equation}
that exploits the knowledge of the position of the first $L$ delta peaks of the distribution $P(m)$ computed solving the underlying equations for $\{m_i\}$ or $m_{ij}$ on small structures. This approach will be developed in a future work.

\section{Comparison between different optimization methods}\label{sec6}

\begin{figure}
\includegraphics[width=0.9\columnwidth]{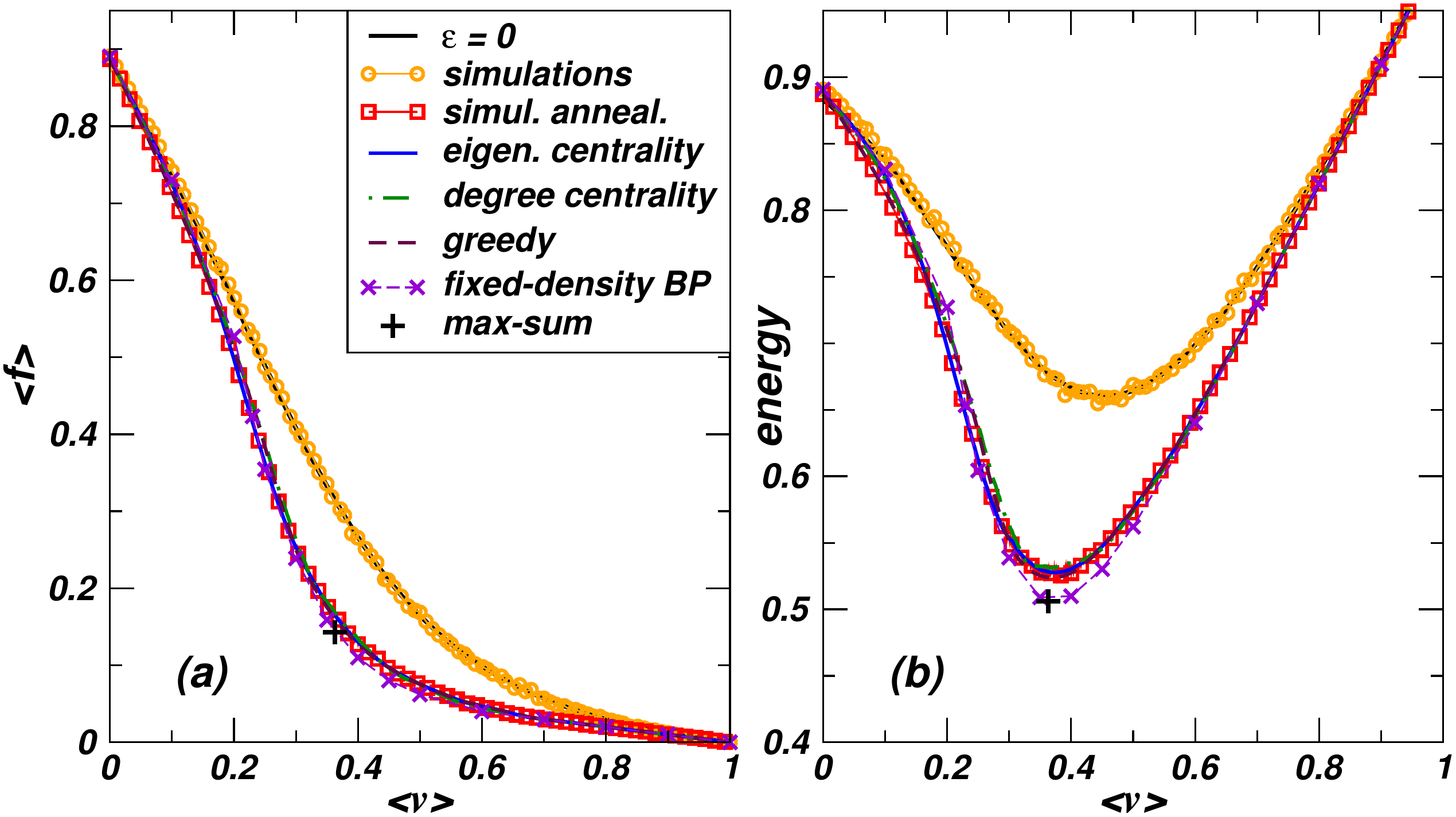}
\caption{Comparison between several immunization methods on a RRG with $N=10^3$ nodes and degree $K=4$ for the SIR model (with $q=0.1$, $p=0.5$): (a) average fraction of infected nodes $\langle f\rangle$ vs. average fraction of immunized nodes $\langle v\rangle$, (b) energy (for $\mu=1.0$) vs. average fraction of immunized nodes $\langle v\rangle$. Reported results include: Simulated annealing (red squares), eigenvectors centrality (blue full line), degree centrality (green dot-dashed line), greedy (maroon dashed line) and the best obtained solving fixed-density BP equations (violet crosses) and MS (black vertical cross). Random immunization is also reported (black line for the BP results with $\epsilon = 0$ and yellow circles for the results of direct stochastic simulations). 
\label{fig-RRGcompare}}
\end{figure}

A comparison between different optimization methods should take into account both the performances of the optimization and the efficiency of the algorithms.
Random regular graphs with uniform parameters are not a particularly good setup to compare different methods because all nodes are equally important. This is visible in Fig.\ref{fig-RRGcompare}a which displays, for the SIR model, the curve $\langle f\rangle$ as function of $\langle v\rangle$ obtained by several different immunization algorithms: a recalculated degree centrality (green dot-dashed line), recalculated eigenvector centrality (blue line), greedy algorithm (maroon dashed line) and density-constrained simulated annealing (red squares). The results on a RRG with $N=10^3$ are compared with the best optimization predicted using density-constrained BP equations on infinitely large RRG (violet crosses). 
For the same optimization problem, the energy as function of the density of immunized nodes is shown in Fig.\ref{fig-RRGcompare}b (we set $\mu=\epsilon=1.0$). Greedy and topologically-based algorithms perform very similarly, and approximately the same result is obtained with fixed-density simulated annealing, even in case of rather slow annealing schedules with up to $10^5$ steps between $\beta=0.1$ and $\beta=10^3$. The BP prediction suggests that (at least for infinitely large graphs) slightly lower energy values could be reached for intermediate values of $\langle v\rangle$. Remarkably, using just $100$ bins, the MS algorithm was able to find an immunization set at such lower energies (black vertical cross), that is expected to be the optimal one. We note that the plain MS equations do not always converge. To overcome this difficulty, we employed the reinforcement technique~\cite{bayati_statistical_2008, bailly-bechet_finding_2011, altarelli_stochastic_2011, altarelli_stochastic_2011-1}. We also ran simulated annealing with $>10^6$ steps and without density constraints, and we found the same optimal point. Notice that, although the computational time of MC based methods is comparable with that of MS for graphs of few thousands of nodes, simulating annealing becomes unfeasible for large-scale networks. 

\begin{figure}
\includegraphics[width=0.49\columnwidth]{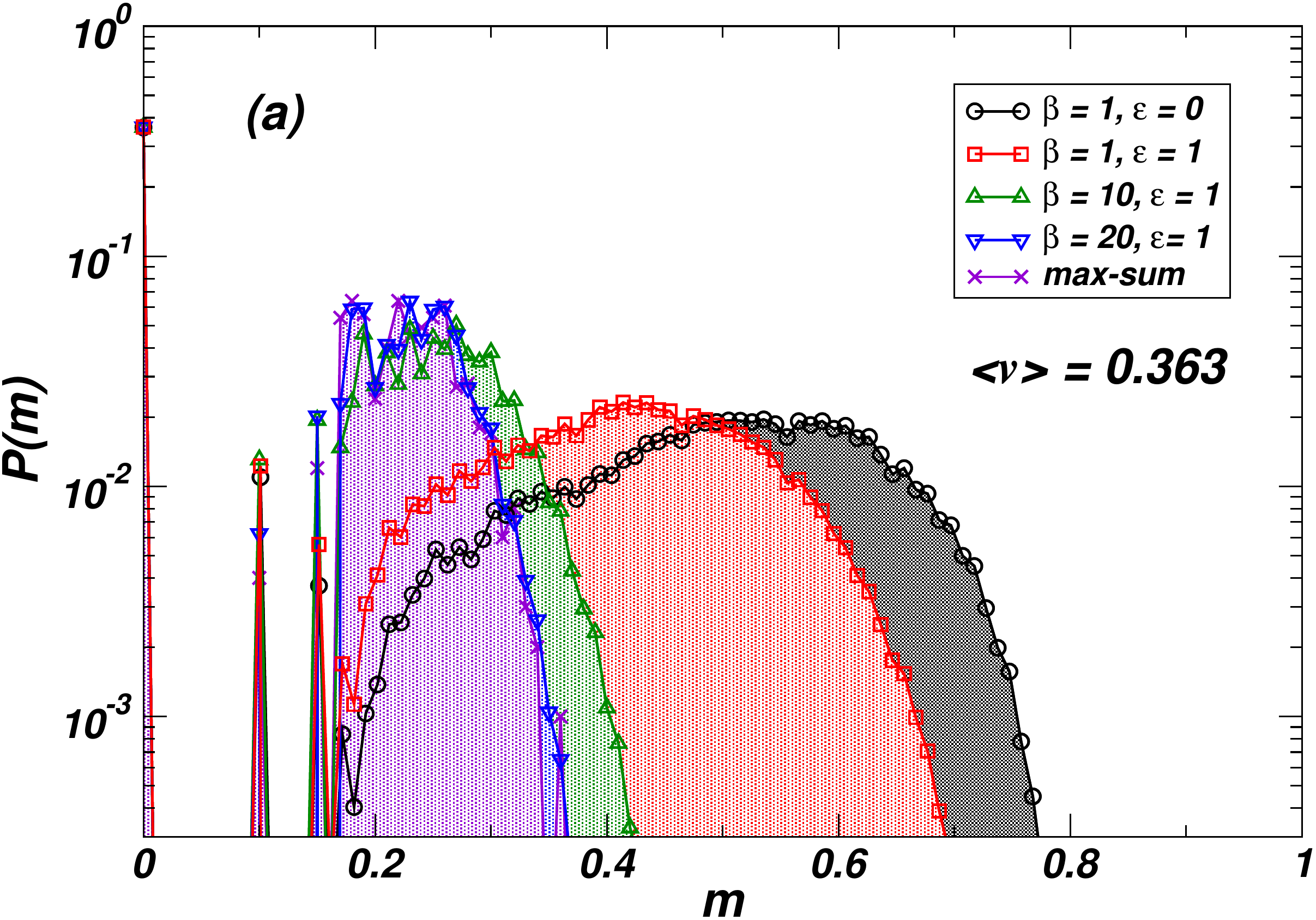}
\includegraphics[width=0.49\columnwidth]{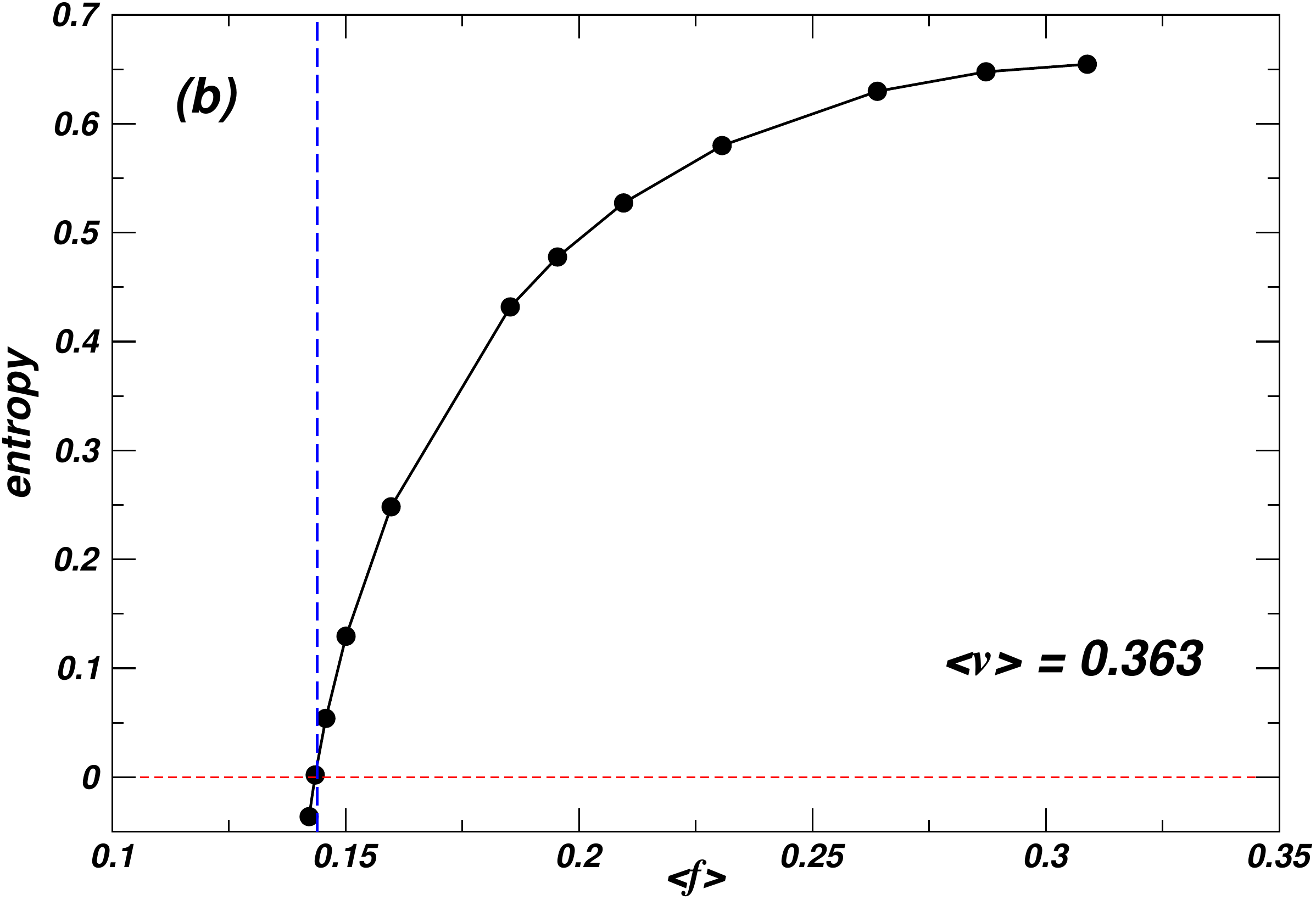}
\caption{(a) Distribution $P(m)$ of the probability $m$ that a node got infected during the SIR process on random regular graphs of degree $K=4$, $q=0.1$, $p=0.5$, $\langle v\rangle= 0.6$ and different values of $\beta$ and $\epsilon$. (b) Entropy of immunization sets of density $\langle v\rangle=0.363$ as function of $\langle f\rangle$ (obtained in creasing $\beta$ for $\epsilon=1$).
\label{fig-optPE-SIR363}}
\end{figure}

We also analyzed some epidemic properties associated with the immunization set found by MS and compared them to those expected solving the BP equations in the single-link approximation at the same density $\langle v\rangle = 0.363$. Figure \ref{fig-optPE-SIR363}a displays the behavior of $P(m)$ computed using BP for different values of $\epsilon$ and $\beta$, and the same quantity obtained solving \eqref{sir-st2}-\eqref{sir-st3} for the configuration of immunized nodes obtained using the MS algorithm (crosses). The accord between the latter and the BP results for large $\beta$ values is very good.   
For the same choice of parameters, Figure \ref{fig-optPE-SIR363}b shows the entropy as function of the density of infected nodes $\langle f \rangle$ (obtained performing BP for increasing values of $\beta$). The entropy decreases monotonically and vanishes at $\langle f \rangle \approx 0.143$, that indicates the minimum level of infection attainable at that density of immunized nodes. The vertical line marking the density of immunized nodes found in the solution obtained using the MS algorithm on a graph of $N=10^3$ nodes perfectly matches this lower bound. 

\begin{figure}
\includegraphics[width=0.9\columnwidth]{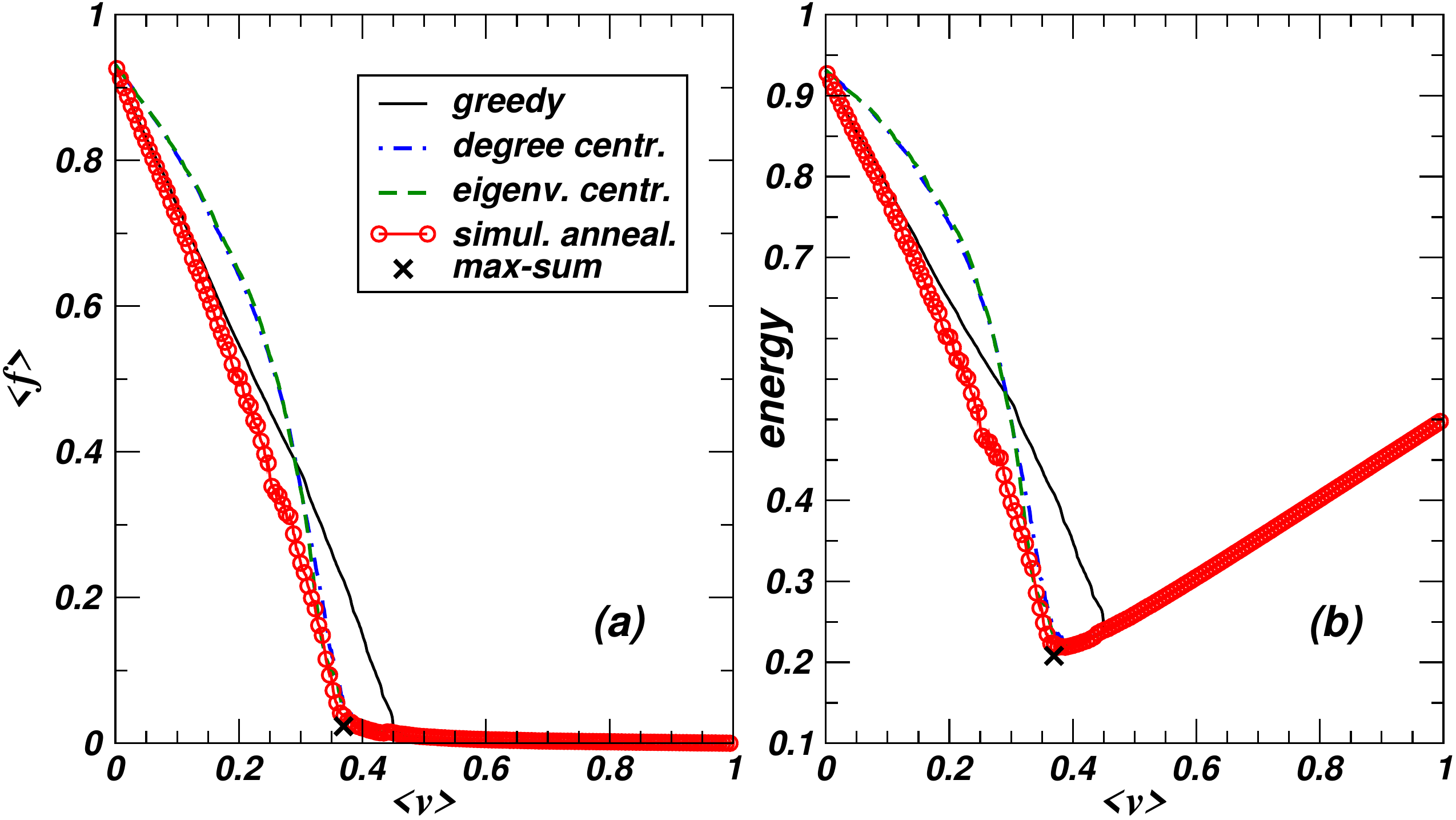}
\caption{Comparison between the optimization of immunization sets for the SIR model (with $q=0.1$, $p=0.5$) on a friendship network of $N = 685$ individuals: (a) average fraction of infected nodes $\langle f\rangle$ vs. average fraction of immunized nodes $\langle v\rangle$, (b) energy (for $\mu=0.5$) vs. average fraction of immunized nodes $\langle v\rangle$. We report results obtained using: an energy-based greedy algorithm (black full line), fixed-density simulated annealing (red circles), degree centrality (blue dot-dashed line) and eigenvector centrality (green dashed line), MS (black cross).
\label{fig-school1SIR}}
\end{figure}

We performed the comparison between different optimization algorithms on a real-world network with non-homogeneous topology. The chosen network, a fully symmetric version of a friendship network among 685 students of a school in the U.S. \cite{moody_peer_2001}, is of moderate size in order to make possible a direct comparison of the different methods with a simulated annealing with sufficiently slow annealing schedule. Figure \ref{fig-school1SIR} shows that, increasing the density of immunized nodes, greedy methods rapidly depart from the optimal curve obtained with density-constrained simulated annealing. On the contrary, heuristic algorithms based of topological metrics performs quite well in finding low energy immunization sets, possibly because of the heterogeneous topological structure of the graph. The optimal point found using the MS algorithm is slightly better than those obtained with all other methods. 

\begin{figure}
\includegraphics[width=0.8\columnwidth]{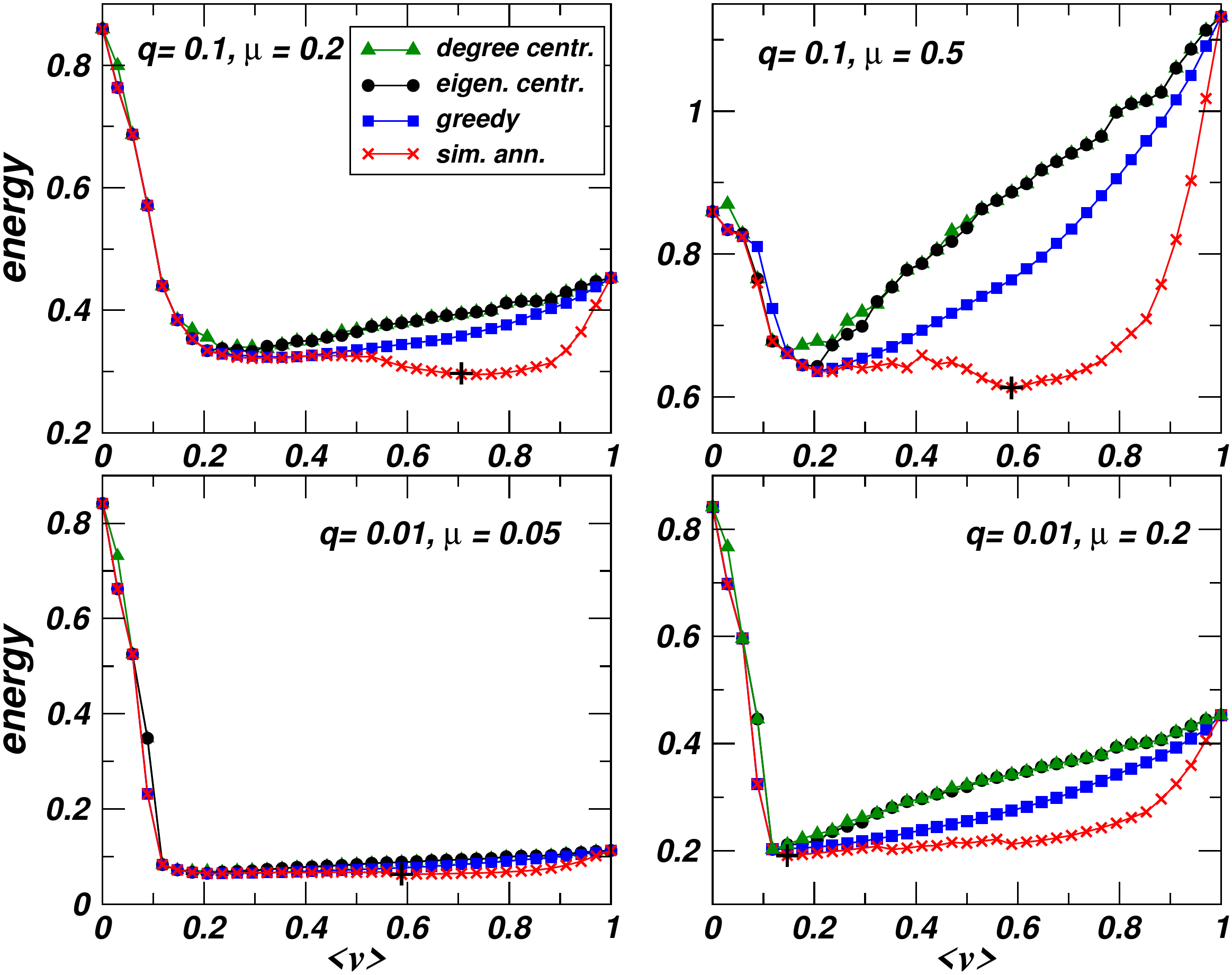}
\caption{Energy as function of $\langle v\rangle$ for immunization sets obtained using degree (green triangles) and eigenvector centrality (black circles), the greedy algorithm (red squares), and fixed-density simulated annealing (blue crosses), for the SIR model on Zachary's karate club network of $N = 34$ individuals. The MS results are indicated by black vertical crosses. The immunization cost was assumed to be half of the degree of a node.
\label{fig-karateSIR}}
\end{figure}

When inhomogeneous costs are added, the algorithms based on topological metrics are ruled out. Although they might work very well for some values of $\mu$ (the parameter governing the tradeoff between terms in the energy), they do not take into account the contribution to the energy associated with the additional parameters. Energy-based methods are more flexible as they adapt to variations of all parameters appearing in the energy. The greedy algorithm is however limited by the incremental procedure, that finds local minima of the energy with a rather small density of immunized nodes. Adding further nodes to the immunization set, the results tend to worsen. On the contrary, simulated annealing and the MS algorithm are able to find immunization sets with a larger number of immunized nodes but overall less expensive. This phenomenon is observed in the data reported in Figure \ref{fig-karateSIR}, where we analyzed immunization strategies for the SIR model on the famous Zachary's karate club network of $N=34$ individuals. On such a small network, the immunization sets obtained by simulated annealing (performed with $10^5$ MC steps between $\beta=0.1$ and $\beta=10^4$) are likely to correspond to the global energy minima (for each value of $\langle v\rangle$). In order to generate a rough energy landscape, we considered immunization costs that are correlated with the degree of the nodes ($c_i = k_i/2$, where $k_i$ is the degree of $i$). Figure \ref{fig-karateSIR} shows that both for $q=0.1$ and $q=0.01$, changing the trade-off between the terms of the energy (i.e. increasing $\mu$) the optimal immunization set is found at very different (larger) values of the density $\langle v\rangle$. MS correctly finds the optimal immunization set for all cases under study (black vertical crosses in Fig.\ref{fig-karateSIR}).

\section{Discussion}

In this work, we considered the optimization problem of targeted network immunization against epidemic spreads. The corresponding energy function to be minimized is a tradeoff between the costs of immunizing nodes and the expected extent of the infection. This optimization problem turns out to be computationally worst-case intractable; it was proven NP hard even for very simple stochastic propagation models. For two prototypical models of stochastic epidemic spreading, the SIR and SIS models, we considered mean-field equations that can be used to evaluate the level of infection in the stationary state associated to every configuration of immunized nodes. The original energy function and the problem of finding the optimal immunization set can be (approximately) recast in terms of these mean-field variables in a mixed representation in which the selection of nodes to be immunized is represented by binary variables and the local level of infection (in the stationary state) is expressed by continuous variables in $[0,1]$. In this formulation, constraints and energy terms are local, allowing the application of the cavity method and the development of efficient message-passing algorithms such as BP. 
Our results obtained using BP equations on random regular graphs shed light on the statistical properties of immunization sets, uncovering in which regions of the parameter space, and to what extent, targeted immunization is actually more effective than random immunization.
The zero-temperature limit of these equations gives the MS algorithm, that can be used to find a solution to the optimization problem. We showed, both on synthetic and real networks, that the MS algorithm outperforms several popular immunization methods based on topological metrics and greedy strategies. The solution found using MS is not guarantee to be optimal, therefore we also performed simulated annealing, that is able to reach the optimum, at least for a sufficiently slow (maybe exponential) annealing schedule. For networks of moderate size, we could compare the lowest-energy immunization set found by MC methods with the solution found by MS. The latter was always at least as good as the former, providing an experimental evidence of the validity of the optimization technique. Moreover, unlike MC-based methods, the MS algorithm scales only linearly with the network size. As a drawback, the algorithm scales as $N_B^3$, where $N_B$ is the number of bins necessary to represent the distribution of real values as histograms. We emphasize that the discretization method used in the current implementation is a very naive and straight-forward one, and there are several ways to considerably reduce $N_B$ by adopting more efficient representation of the messages (as explained in Sec.\ref{sec5}). For this reason we expect that message-passing algorithms as the one proposed here could be used to study the immunization problem even on very large networks. This will be the scope of future research.

The results of the comparison between different optimization methods on a variety of networks show that, as long as the
network and the parameters are sufficiently homogeneous, all methods give approximately the same results. In this case, heuristic strategies based on topological metrics, such as degree-centrality or eigenvector centrality could be preferred, as they are very simple and fast. When the energy function includes inhomogeneous costs, simpler heuristics turned out to be sub-optimal in our experiments. Moreover, greedy-based methods, that take into account the correct energy function, are based on a progressive scheme (as immunized nodes are added incrementally) and often fail to reach the ground state whenever the energy landscape is rugged. This is well demonstrated by results obtained on the small, but representative, Zachary's karate club network.

The method presented here could be applied to a number of other optimization problems including, but not restricted to, other epidemic models in discrete or continuous time, provided that similar fixed-point equations are defined for node or edge variables. The control variables in that case could be a set of node-dependent external parameters. Using message-passing techniques it could be possible to select the configuration of external parameters that corresponds to some desired outcome for the global state of the system. This general formulation opens to a wide spectra of applications in problems involving the control of network dynamics.

\appendix

\section{Algorithms}\label{app-algo}
\subsection{Incremental Heuristic Algorithms}

The heuristic algorithms considered in the paper are based on an incremental procedure, by means of which the algorithm adds one by one nodes to the immunization set, each time choosing the node that minimizes some score function. Hence, given a graph $G = (V,E)$ and one of the scoring strategies described below, the immunization algorithm is based on the following iteration:
\begin{itemize}
\item[] for $k=1,\dots, |V|$:
\begin{itemize}
	\item[1.] [score] compute the score vector $\bf{z}$;
	\item[2.] [immunization] choose node index $i^{*}_{k}$ corresponding to the highest component of score $\bf{z}$ and removed from it the graph;
\end{itemize} 
\end{itemize}
This procedure gives an immunization set $\{i^*_1,\dots,i^{*}_{k}\}$ for any desired number $k$ of immunized nodes.

The actual definition of the score vector depends on the heuristic strategy considered. For degree centrality, the score of a node is just the number of non-immunized neighbors. 
For the eigenvector centrality, the score of a node is recursively a function of the scores of neighbors, 
\begin{equation}\label{eigen}
z_i = \frac{1}{\lambda}\sum_{j \in \partial i} z_j  
\end{equation}
where $\lambda$ is a properly defined constant. The score vector $\bf{z}$ satisfies the eigenvector equation $A \bf{z} = \lambda \bf{z}$, where $A$ is adjacency matrix, such that $a_{ij} = 1$ if there is an edge between nodes $i$ and $j$ and $a_{ij} = 0$ otherwise. Under the condition that $\bf{z}> 0$ and the graph is connected, the constant $\lambda$ corresponds to the greatest eigenvalue of the adjacency matrix (Perron-Frobenius Theorem). Hence, the score vector $\bf{z}$ can be computed by iteration from a homogeneous initial condition $z_i^0=1$, $\forall i\in V$ using the Power Method, i.e. defining 
\begin{equation}
z_i^{t+1} = \frac{1}{\lambda_t}\sum_{j=1}^{N} a_{ij} z_j^t  
\end{equation}
where $\lambda_t$ is an appropriately defined normalization constant (recomputed at each time step). If the iteration converges, it gives the eigenvector centrality of the nodes.

In the greedy algorithm implemented in the present paper, the score of a node $i$ is equal to the variation in energy, computed from \eqref{energy}, associated with the addition of $i$ to the immunization set. Given ${\bf s}=(s_1, \dots, s_{i-1}, 0, s_{i+1}, \dots, s_N)$ and ${\bf s}'=(s_1, \dots, s_{i-1}, 1, s_{i+1}, \dots, s_N)$, we define 
\begin{equation}
z_i = \mathscr{E}(\bf{s}) - \mathscr{E}(\bf{s}').
\end{equation}   

\subsection{Simulated Annealing}
The space of $2^N$ binary configurations corresponding to immunization sets can be sampled using a Montecarlo algorithm, that at large inverse temperature converges towards the the minima of the energy function $\mathscr{E}$. In practice, starting from a randomly selected binary configuration, the convergence to a global minimum can be achieved only using an annealing schedule that guarantees a sufficiently slow decrease in temperature.  
Given a randomly chosen initial condition ${\bf s} = (s_1, \dots ,s_N)$, and an initial value $\beta_I$ for the inverse temperature $\beta$, the adopted annealing schedule reaches a final value $\beta_F$ in a number $M$ of proposed single-spin flip. Unfortunately, the number $M$ of steps required to reach the minimum of the energy at large $\beta$ often scales exponentially with the system size $N$.

In summary, the algorithm works as follows:
\begin{itemize}
\item[0.] choose an initial condition  ${\bf s}$, set $\beta \leftarrow \beta_I$;
\item[1.] randomly select a node $i$ to be flipped; 
\item[2.] given ${\bf s}=(s_1, \dots, s_{i-1}, s_{i}, s_{i+1}, \dots, s_N)$ and ${\bf s}'=(s_1, \dots, s_{i-1}, 1-s_{i}, s_{i+1}, \dots, s_N)$ compute the variation of energy $\Delta \mathscr{E} = \mathscr{E}(\bf{s}') - \mathscr{E}(\bf{s}) $;
\item[3.] accept the move ${\bf s} \leftarrow {\bf s}'$ with probability $e^{-\beta\Delta\mathscr{E}}$;
\item[4.] set $\beta \leftarrow \beta + \delta\beta$
\item[5.] if $\beta < \beta_F$, then go to point 1.  
\end{itemize}
In our simulations we tested different experimental setups, using both linear schedule with $\delta\beta = (\beta_F - \beta_I)/M$ and a faster exponential one, in which $\delta\beta = \beta \left[\left(\beta_F/\beta_I\right)^{1/M}-1\right]$. We usually considered $\beta_I < 1$ and $\beta_F$ between $10^3$ and $10^4$ with a number of single-spin proposed moves $M$ between $10^5$ and $10^6$. 

In the fixed-density simulated annealing algorithm, we considered only moves that do no change the number of immunized nodes, such as immunization exchange moves. In this case, the overall algorithm consists in repeating the annealing schedule for any possible value of the number of immunized nodes from $0$ to $N$.

\end{document}